\documentstyle[prd,aps,eqsecnum,twocolumn,floats]{revtex}

\def\beq{\begin{equation}}
\def\eeq{\end{equation}}
\def\beqa{\begin{eqnarray}}
\def\eeqa{\end{eqnarray}}
\def\bfig{\begin{figure}}
\def\efig{\end{figure}}

\input{psfig}

\begin{document}
\fnsymbol{footnote}
\draft
\wideabs{

\title{{\it R-}Modes in Superfluid Neutron Stars}

\author{Lee Lindblom}
\address{Theoretical Astrophysics 130-33,
         California Institute of Technology,
         Pasadena, CA 91125}
\author{Gregory Mendell}
\address{Department of Physics and Astronomy,
         University of Wyoming,
         Laramie, WY 82071}

\date{September 27, 1999}
\maketitle
\begin{abstract}
The analogs of $r$-modes in superfluid neutron stars are studied here.
These modes, which are governed primarily by the Coriolis force, are
identical to their ordinary-fluid counterparts at the lowest order in
the small angular-velocity expansion used here.  The equations that
determine the next order terms are derived and solved numerically for
fairly realistic superfluid neutron-star models.  The damping of these
modes by superfluid ``mutual friction'' (which vanishes at the lowest
order in this expansion) is found to have a characteristic time-scale
of about $10^4$ s for the $m=2$ $r$-mode in a ``typical'' superfluid
neutron-star model.  This time-scale is far too long to allow mutual
friction to suppress the recently discovered gravitational radiation
driven instability in the $r$-modes.  However, the strength of the
mutual friction damping depends very sensitively on the details of the
neutron-star core superfluid.  A small fraction of the presently
acceptable range of superfluid models have characteristic mutual
friction damping times that are short enough (i.e. shorter than about
5~s) to suppress the gravitational radiation driven instability
completely.

\end{abstract}
}

\section{Introduction}
\label{sectionI}

Recently Andersson~\cite{andersson} and Friedman and
Morsink\cite{fried-morsink} showed that the $r$-modes in {\it all}
rotating stars would be driven unstable by the emission of
gravitational radiation in the absence of internal fluid
dissipation.  Subsequent analysis by Lindblom, Owen, and
Morsink~\cite{lom} and then by Andersson, Kokkotas, and
Schutz~\cite{aks} showed that internal fluid dissipation in hot young
neutron stars is insufficient to suppress this gravitational radiation
driven instability.  Thus neutron stars that are formed rapidly
rotating are expected to spin down within about one year to a
relatively small angular velocity (about 5--10\% of the maximum) by
the emission of gravitational radiation.  Owen et al.~\cite{owen-etal}
constructed rough models of this spindown process, and concluded that
the gravitational radiation from these spindown events might be
observable by the second-generation LIGO gravitational wave detectors.

The purpose of this paper is to investigate the behavior of this
gravitational-radiation instability in the $r$-modes of older colder
neutron stars~\cite{akst,bildsten,levin}.  Here the physics is more
complicated and there are interesting observational constraints.  The
existence of the two 1.6 ms pulsars~\cite{backer-etc}, and numerous
examples of somewhat more slowly rotating neutron stars in low mass
x-ray binaries (LMXBs)~\cite{backer} show that some neutron stars are
in fact rapidly rotating and stable.  Unfortunately uncertainty in the
neutron-star equation of state means that the minimum rotation periods
for neutron stars are not presently well known, and so we can not say
exactly how rapid (in a dynamically meaningful sense) these rotations
really are.  Values for the minimum rotation periods (when mass
shedding first occurs) of 1.4~$M_\odot$ models range from 0.5 ms to
about 1.4 ms, depending on the equation of state~\cite{cook-etal}.
Any value in this range, however, implies that the 1.6 ms pulsars are
rotating much more rapidly than is consistent with our present
understanding of the gravitational wave spindown due to the $r$-mode
instability in hot young neutron stars.  It is widely believed,
however, that these 1.6 ms objects are old cold recycled
pulsars~\cite{phinney}, having been spun up by accretion long after
their initial cool-down.  These neutron stars are expected to have
superfluid cores, and hence the fluid dynamics and dissipation
mechanisms that govern their $r$-modes are entirely different from
those studied to
date~\cite{andersson,fried-morsink,lom,aks,owen-etal}.  The purpose of
this paper is to develop the tools needed to study the $r$-modes in
superfluid neutron stars.  The challenge is to understand how the
$r$-mode instability is suppressed in the 1.6 ms pulsars in
particular, and the more numerous 3 ms objects in LMXBs more
generally.

The superfluid dissipation mechanism called ``mutual friction'' seems
a likely candidate to provide the needed stability for the $r$-modes
in old cold neutron stars.  Mutual friction arises from the scattering
of electrons off the magnetic fields entrapped in the cores of the
superfluid neutron vortices~\cite{alpar-langer-sauls,alpar-sauls}, and
is known to play an important role in other aspects of the dynamics of
superfluid neutron stars.  Lindblom and Mendell~\cite{lind-mendell95}
show, for example, that mutual friction completely suppresses the
gravitational radiation driven instability in the $f$-modes of
rotating neutron stars.  Our results here for the $r$-modes present a
more ambiguous picture.  We find in Sec.~\ref{sectionVI} that the
characteristic damping time for the $r$-modes due to mutual friction
is about $10^4$~s, for a typical model of the neutron-star core
superfluid.  This timescale is far too long to have any appreciable
effect on the $r$-mode instability in these stars.  However, we also
find that the mutual-friction damping time is extremely sensitive to
the parameters that define the core superfluid.  Within the presently
acceptable range of the parameters examined here, about 1\% have
mutual friction damping times so short (i.e. shorter than about 5 s)
that the $r$-mode instability is suppressed completely.  A somewhat
larger fraction of these parameters in the acceptable range, about 3\%, have
damping times short enough (i.e. shorter than about 58 s) that mutual
friction suppresses the instability in some sufficiently warm and
sufficiently slowly rotating neutron stars.  Thus we conclude that an
appropriately fine-tuned superfluid dynamics could provide the needed
stability for the $r$-modes in old cold neutron stars through mutual
friction.  However given the small fraction of superfluid models that
provide the needed stability, other damping mechanisms need to be
considered (e.g. solid crust effects, strange quark matter, magnetic
fields, ...).  

Regular shear viscosity is another mechanism that could play a role in
damping the $r$-modes of superfluid neutron stars.  We analyze this
damping mechanism for the superfluid $r$-modes in Sec.~\ref{sectionVI}
and find a characteristic damping time of about $10^8(T/10^9 {\rm
K})^2$~s, where $T$ is the temperature of the neutron-star core.  This
time-scale is short enough to suppress the $r$-mode instability in
stars cooler than about $10^6$~K. Neutron stars are spun up in the
usual picture during an LMXB phase, in which the core temperature is
expected to exceed $10^8$~K~\cite{brown}.  This temperature is too hot
to allow shear viscosity to provide the needed stability even for the
3~ms neutron stars observed in these systems.  Levin~\cite{levin} has
shown that when a neutron star is spun up to the point where stability
of the $r$-modes is lost, the star heats up and then spins down to a
very small angular velocity in a few months by emitting gravitational
radiation.  Thus, some robust internal fluid dissipation mechanism
must be identified to explain the stability of the $r$-modes in the
observed LMXB systems.  If mutual friction is the only mechanism
capable of providing the needed stability, then this fact would place
interesting constraints on the parameters of the neutron-star core
superfluid.  An alternate possibility in the case of the 1.6~ms
pulsars would be a mechanism for spinning up these stars without
raising their core temperatures above about $10^7$~K (e.g. by
accretion at very low rates).

In Sec.~\ref{sectionII} we review the basic hydrodynamics of
neutron-star core superfluids.  We outline in Sec.~\ref{sectionIII}
the derivation of the equations that govern the normal modes of a
superfluid neutron star from this hydrodynamic theory.  In
Sec.~\ref{sectionIV} we take the small angular velocity expansions of
these equations that are needed to study the $r$-modes.  In
Sec.~\ref{sectionV} we present our numerical solutions for the
$r$-modes of rotating superfluid neutron stars, up to the second order
in the small angular velocity expansion.  The effects of superfluid
mutual friction and shear viscosity on these $r$-modes are evaluated
in Sec.~\ref{sectionVI}.  The equations that determine the superfluid
pulsations are expressed in spherical coordinates in the Appendix.

\section{Superfluid Hydrodynamics in Neutron-Star Matter}
\label{sectionII}

When the core temperature of a neutron star drops below about $10^9$
K, a phase transition to a superconducting-superfluid state is
expected to occur~\cite{baym-patheck,pines-alpar,sauls,epstein}. The
neutrons in the core are expected to form ${}^3P_2$ Cooper pairs and
the protons to form ${}^1S_0$ pairs.  The purpose of this section is
to review briefly the equations that describe the behavior of this
complicated superconducting-superfluid mixture on the macroscopic
scales needed here to describe the $r$-modes.  

Let $\vec{v}_n$ denote the velocity of the neutron superfluid, and
$\vec{v}_p$ the velocity of the proton superfluid.  On small scales
these superfluid velocities, $\vec{v}_n$ and $\vec{v}_p$, are related
by the London equations to the phases of the complex order parameters,
$S_n$ and $S_p$, that describe the neutron and proton condensates;
thus, $\vec{v}_n = (\hbar/2m_n) \vec{\nabla}S_n$ and $\vec{v}_p
=(\hbar/2m_p) \vec{\nabla}S_p - (e/m_pc)\vec{A}$, where $\vec{A}$ is
the electromagnetic vector potential.  These equations imply that
vorticity and magnetic fields in this material are confined to
vortices and flux tubes of microscopic dimension.  In a typical
neutron star the spacing between these neutron vortices is expected to
be of order $10^{-3}$ cm, while the spacing between magnetic
flux-tubes is expected to be about $10^{-10}$ cm~\cite{sauls}.  Our
interest here is the very large-scale motions of this material
associated with the low-order $r$-modes.  Thus, it is appropriate to
consider all physical quantities, such as $\vec{v}_n$ and $\vec{v}_p$,
to be averaged over many vortices.  The procedure for making this
average is described more fully by Bekarevich and
Khalatnikov\cite{bekarevich-khalatnikov1961}, Baym and
Chandler~\cite{baym-chandler1983}, Sonin~\cite{sonin1987}, Mendell and
Lindblom~\cite{mendell-lindblom1991}, and
Mendell~\cite{mendell1991,mendell98}.  Throughout the remainder of
this paper, all quantities are assumed to be so averaged.

One of the interesting and unusual features of the neutron-star core
superfluid is the so called ``drag
effect''~\cite{andreev-bashkin,alpar-langer-sauls,sauls} or
``entrainment effect''~\cite{carter-langlois1998}.  This effect is
caused by the fact that the conserved particle currents are not simply
proportional to the superfluid velocities.  Instead these conserved
currents are linear combinations of $\vec{v}_n$ and $\vec{v}_p$:
$\rho_{nn}\vec{v}_n + \rho_{np}\vec{v}_p$ for the neutron current, and
$\rho_{pp}\vec{v}_p + \rho_{np}\vec{v}_n$ for the proton current.
Thus a given neutron superfluid flow $\vec{v}_n$ is accompanied by a
certain (small) current of protons, and vice versa.  The mass-density
matrix elements $\rho_{nn}$, $\rho_{pp}$, and $\rho_{np}$ are
determined by the micro-physics of the many-body strong-interactions
that occur between the neutrons and protons.  This entrainment effect
plays a crucial role in mutual friction (perhaps the most important
dissipation mechanism in this material) which we discuss in more
detail in Sec.~\ref{sectionVI}.  Unfortunately these mass-matrix
elements are not well determined at the present time.  These
quantities are constrained by Galilean invariance: $\rho_{nn} = \rho_n
- \rho_{np}$ and $\rho_{pp} = \rho_p - \rho_{np}$, where $\rho_n$ and
$\rho_p$ are the neutron and proton mass densities.  But, the
independent element $\rho_{np}$ must be determined directly from the
micro-physics.  We find it convenient to re-express $\rho_{np}$ in
terms of the dimensionless entrainment parameter $\epsilon$:

\beq
\rho_{np}=-\epsilon\rho_n.\label{2.1}
\eeq

\noindent Borumand, Joynt, and Klu\'zniak~\cite{bjk} estimate that
$\epsilon\approx 0.04$, and that its value is known at present only to
within about a factor of two.  Given this uncertainty we explore the
properties of the $r$-modes over the expected range of superfluid
models with $0.02\leq \epsilon \leq 0.06$.

The material in the core of a neutron star is a complicated mixture of
neutrons, protons, electrons, muons, etc.  While the general equations
that describe the dynamics of this kind of charged
superconducting-superfluid mixture have been
studied~\cite{mendell-lindblom1991}, these general equations are
considerably more complicated than are needed here.  Our present
interest is the dynamics of the superfluid analogs of the $r$-modes;
thus we are interested in dynamics having length-scales comparable to
the size, and time-scales comparable to the rotation period, of the
star.  Under these conditions the dynamics of the core superfluid
material simplifies considerably.  On time-scales longer than the
plasma time-scale (about $10^{-21}$ s) and the cyclotron time-scale
(about $10^{-15}$ s), for example, this material is well described by
the magnetohydrodynamic limit of the exact equations~\cite{mendell98}.
In this limit the electrons and muons maintain exact charge neutrality
with the protons.  Similarly the electrons and muons are forced by
scattering to move together as a single fluid on time-scales longer
than about $10^{-9}$ s~\cite{sauls}.  Further, for dynamics on the
time-scales of interest here, the bulk electrical currents are
extremely small~\cite{mendell98}.  Thus it is appropriate to simplify
further and require the charged species to move together without
generating any electrical current.  The dynamical degrees of freedom of
this material are reduced therefore to a pair of velocity vector
fields---one for the neutrons and one for the protons---and a
corresponding pair of thermodynamic scalar densities.

In general there are forces in the complete dynamical equations (even
for this reduced system) that describe the interactions between the
smoothed superfluid flow and the sheaf of
vortices~\cite{mendell-lindblom1991}.  These additional forces are
negligible for fluid motions with time-scales comparable to the
$r$-modes~\cite{mendell1991,mendell98}, and we neglect them here.
And finally, in
general the dynamics would also include a ``normal'' component of this
superfluid material; however, again we simplify by assuming that the
temperature is well below the superfluid transition and ignore these
additional dynamical degrees of freedom.

Our study is directed toward an exploration of the superfluid analogs
of the $r$-modes.  We are primarily interested therefore in examining
the equations that describe the evolution of small departures from a
uniformly rotating equilibrium neutron star.  The dynamics of the
neutron-star core superfluid is described by two velocity vectors and
two thermodynamic scalar fields.  It will be convenient to express the
equations for these velocity fields in terms of $\delta\vec{v}$ and
$\delta\vec{w}$: the average and relative velocities of the core
superfluids.  These quantities are defined as

\beq
\rho\, \delta \vec{v}=\rho_n\delta\vec{v}_n +\rho_p\delta\vec{v}_p,
\label{2.2}
\eeq

\noindent and
\beq
\delta\vec{w}= \delta\vec{v}_p - \delta\vec{v}_n.
\label{2.3}
\eeq

\noindent We use the prefix $\delta$ to denote a small (Eulerian)
perturbation away from the equilibrium value of a quantity; while,
quantities without prefix, such as $\rho=\rho_n+\rho_p$, denote the
equilibrium values.  The superfluid velocity fields $\delta\vec{v}_n$
and $\delta\vec{v}_p$ are easily determined from Eqs.~(\ref{2.2}) and
(\ref{2.3}) once $\delta\vec{v}$ and $\delta\vec{w}$ are known.
Similarly the velocity field of the electrons, $\delta \vec{v}_e$, can
can be expressed in terms of these quantities from the condition that
there is no electrical current~\cite{mendell1991}:

\beq \delta\vec{v}_e = {\rho_{pp}\over \rho_p}\delta \vec{v}_p
+{\rho_{np}\over\rho_p}\delta\vec{v}_n = \delta \vec{v} +
{\rho_n\over\rho} \left(1+\epsilon
{\rho\over\rho_p}\right)\delta\vec{w}.
\label{2.4}
\eeq

\noindent We note that for simplicity in these equations we have
ignored terms of order $m_e/m_p$, the ratio of electron to proton
mass.  And for simplicity in this discussion, we have also ignored the
presence of muons.  A more complete discussion including the
contributions of the muons is given by Mendell~\cite{mendell1991}.

The equations for the velocity fields $\delta\vec{v}$ and
$\delta\vec{w}$ are obtained by perturbing the full system of
superconducting-superfluid evolution equations, subject to the
assumptions described above.  These equations, when expressed in terms
of $\delta\vec{v}$ and $\delta\vec{w}$, have the remarkably simple
forms~\cite{mendell-lind}:

\beqa
\partial_t\delta v^a + v^b\nabla_b\delta v^a &&+ \delta v^b\nabla_b v^a
= \nonumber\\
&&-\nabla^a\delta U + {1\over\rho^2}\left({\partial\rho\over\partial\beta}
\right)_p\delta\beta\,\nabla^a p,
\label{2.5}
\eeqa

\beqa
\partial_t\delta w^a + v^b\nabla_b\delta w^a 
+(2\gamma-1)\delta w^b\nabla_b v^a=-\nabla^a\delta\beta.
\label{2.6}
\eeqa

\noindent The perturbed scalar $\delta U$ that appears on the right
side of Eq.~(\ref{2.5}) is defined by

\beq
\delta U = {\delta p\over \rho}-\delta\Phi,\label{2.7}
\eeq

\noindent where $\delta p$ is the perturbed pressure, and $\delta\Phi$
the perturbed gravitational potential.  The potential $\delta\beta$
that appears on the right sides of Eqs.~(\ref{2.5}) and (\ref{2.6})
measures the degree to which the perturbed fluid departs from
$\beta$-equilibrium.  The thermodynamic function $\beta$ is related to
the chemical potentials (per unit mass) of the neutrons $\mu_n$,
protons $\mu_p$, and electrons $\mu_e$ by

\beq
\beta = \mu_p - \mu_n +{m_e\over m_p}\mu_e.\label{2.8}
\eeq

\noindent  The quantity $\beta$ vanishes in the equilibrium
state.  (For simplicity we again neglect terms of order $m_e/m_p$.)
Finally, the velocity field $v^a$ in Eqs.~(\ref{2.5}) and (\ref{2.6})
represents the uniform rotation of the equilibrium star, and
the dimensionless quantity $\gamma$ that appears in Eq.~(\ref{2.6}) is
related to the determinant of the superfluid mass-density matrix:
$\gamma=(\rho_{nn}\rho_{pp}- \rho_{np}^2) /\rho_n \rho_p$.

The three perturbed scalar fields, $\delta U$, $\delta\Phi$, and
$\delta\beta$, that appear on the right sides of Eqs.~(\ref{2.5}) and
(\ref{2.6}) can be used to determine all of the other scalars of
interest in this problem; for example

\beq
\delta\rho = \rho\left({\partial\rho\over\partial p}\right)_\beta
(\delta U +\delta\Phi)+
\left({\partial\rho\over\partial\beta}\right)_p\delta\beta,
\label{2.9}
\eeq

\noindent and $\delta\rho_p$ (and thence $\delta\rho_n$ as well) can be
determined from

\beqa
\delta\beta =
&&\biggl[
\left({\partial\mu_p\over\partial\rho_n}\right)_{\rho_p}
- \left({\partial\mu_n\over\partial\rho_n}\right)_{\rho_p}
\biggr](\delta\rho-\delta\rho_p)
\nonumber\\
&&+\biggl[\left({\partial\mu_p\over\partial\rho_p}\right)_{\rho_n}
- \left({\partial\mu_n\over\partial\rho_p}\right)_{\rho_n}
+ {m_e^2\over m_p^2}{d\mu_e\over d\rho_e}
\biggr]\delta\rho_p.
\label{2.10}
\eeqa

\noindent It is straightforward then to transform the mass
conservation laws (for neutrons and protons) into forms depending only
on $\delta U$, $\delta\Phi$, and $\delta\beta$~\cite{mendell-lind}:

\beq
\partial_t\delta\rho + v^a\nabla_a\delta\rho + \nabla_a(\rho\delta v^a)=0,
\label{2.11}
\eeq

\beqa
&&(\partial_t + v^a\nabla_a)\biggl[
\left({\partial\rho\over\partial\beta}\right)_p(\delta U+\delta\Phi)
+{\rho_n^2\over\rho}{\partial\over\partial\beta}
\left({\rho_p\over\rho_n}\right)_p\delta\beta\biggr]\nonumber\\
&&\qquad+{1\over\rho}\left({\partial\rho\over\partial\beta}\right)_p
\delta v^a\nabla_a p 
+\nabla_a\left(\tilde{\rho}\,\delta w^a\right)=0.
\label{2.12}
\eeqa

\noindent The quantity $\tilde{\rho}$ that appears in Eq.~(\ref{2.12})
is defined as $\tilde{\rho}=(\rho_{nn}\rho_{pp}-\rho_{np}^2)/\rho=
\rho_n\rho_p\gamma/\rho$.  These Eqs.~(\ref{2.5}), (\ref{2.6}),
(\ref{2.11}), and (\ref{2.12}), together with the perturbed
gravitational potential equation,

\beq
\nabla^a\nabla_a\delta\Phi = -4\pi G\delta\rho.
\label{2.13}
\eeq

\noindent determine the evolution of the material in the superfluid
core of a neutron star in the long length-scale, long time-scale, and
low temperature approximation of interest to us here.

\section{Oscillations of Superfluid Neutron Stars} 
\label{sectionIII}

A superfluid neutron star is a reasonably complicated structure
consisting of a superfluid core surrounded by a solid crust, and
probably a liquid ocean above that.  For the purposes of our analysis
here we use a simplified and idealized representation of this
structure.  We consider a neutron-star model that consists of a
superfluid core (where $\rho>\rho_s$), surrounded by an ordinary
matter envelope (where $\rho<\rho_s$).  For simplicity we treat the
material in this envelope as a perfect fluid. The dynamics of the
material in the core is described by Eqs.~(\ref{2.5}), (\ref{2.6}),
(\ref{2.11}), (\ref{2.12}), and (\ref{2.13}).  And similarly the
material in our idealized envelope is described by Euler's equation,
Eq.~(\ref{2.5}) with $\delta\beta=0$, and Eqs.~(\ref{2.11}) and
(\ref{2.13}).  In this section we show how the modes of such a
rotating superfluid stellar model can be described completely in terms
of the three scalar potentials $\delta U$, $\delta \beta$ and
$\delta\Phi$~\cite{mendell-lind}.  And we derive the
equations that determine these potentials, together with the
appropriate boundary conditions.

We assume here that the time dependence of the perturbation is
$e^{i\omega t}$ and that its azimuthal angular dependence is
$e^{im\varphi}$, where $\omega$ is the frequency of the mode and $m$
is an integer.  The superfluid versions of the Euler equation,
Eqs.~(\ref{2.5}) and (\ref{2.6}), determine the velocities $\delta
v^a$ and $\delta w^a$ in terms of the scalars $\delta U$ and $\delta
\beta$ much as they do in the ordinary-fluid case~\cite{ipser-lind}.
Given the temporal and angular dependence assumed here, Eqs.~({2.5})
and (\ref{2.6}) become linear algebraic equations for $\delta v^a$ and
$\delta w^a$ which can be solved directly:

\beq
\delta v^a = iQ^{ab}\left[\nabla_b\delta U-{1\over \rho^2}
\left({\partial\rho\over \partial\beta}\right)_p\delta \beta\nabla_bp\right]
,\label{3.1}
\eeq

\beq
\delta w^a = i \tilde{Q}^{ab}\nabla_b\delta\beta.\label{3.2}
\eeq

\noindent In these equations $Q^{ab}$ and $\tilde{Q}^{ab}$ are tensors
that depend on the frequency of the mode $\omega$, and the angular
velocity of the equilibrium star $\Omega$.  These tensors are given
by:

\beqa
Q^{ab}=&&{1\over (\omega+m\Omega)^2-4\Omega^2}\nonumber\\
&&\times\Biggl[(\omega+m\Omega)\delta^{ab}-
       {4\Omega^2z^az^b\over \omega+m\Omega} - 2i\Omega\nabla^a\varphi^b
\Biggr],
\label{3.3}
\eeqa

\beqa
\tilde{Q}^{ab} &&= 
{1\over (\omega+m\Omega)^2-4\gamma^2\Omega^2}\nonumber\\
&&\times\Biggl[(\omega+m\Omega)\delta^{ab}-
       {4\gamma^2\Omega^2z^az^b\over \omega+m\Omega} 
- 2i\gamma\Omega\nabla^a\varphi^b
\Biggr].
\label{3.4}
\eeqa

\noindent In Eqs.~(\ref{3.3}) and (\ref{3.4}) $\Omega$ is the angular
velocity of the equilibrium star; the unit vector $z^a$ points along
the rotation axis; $\varphi^a$ is the vector field that generates
rotations about the $z^a$ axis; and $\delta^{ab}$ is the Euclidean
metric tensor (the identity matrix in Cartesian coordinates).

The expressions for the velocity fields in Eqs.~(\ref{3.1}) and
(\ref{3.2}) can be substituted into the mass conservation laws,
Eqs.~(\ref{2.11}) and (\ref{2.12}), to obtain equations for the scalar
fields alone.  In general, the potentials $\delta U$, $\delta\beta$
and $\delta \Phi$ are solutions then of the following system of
partial differential equations~\cite{mendell-lind}:
\vfill\break
\beqa 
\nabla_a\bigl(\rho &&Q^{ab}\nabla_b\delta U\bigr)+(\omega+m\Omega)
\rho\left({\partial\rho\over \partial p}\right)_\beta\delta U=\nonumber \\
&&\nabla_a\left[{1\over\rho}\left({\partial\rho\over\partial\beta}\right)_p
\delta\beta\, Q^{ab}\nabla_bp\right]\nonumber \\
&&-(\omega+m\Omega)
\left[\left({\partial \rho\over \partial\beta}\right)_p\delta\beta
+\rho\left({\partial\rho\over \partial p}\right)_\beta\delta\Phi\right],
\label{3.5} 
\eeqa

\beqa
\nabla_a&&\bigl(\tilde{\rho}\tilde{Q}^{ab}\nabla_b\delta \beta\bigr) 
+(\omega+m\Omega){\rho_n^2\over \rho}{\partial\over \partial\beta}
\left({\rho_p\over \rho_n}\right)_p\delta\beta\nonumber \\
&&-{1\over \rho^3}\left({\partial\rho\over\partial\beta}\right)_p^2
Q^{ab}\nabla_ap\nabla_bp\,\,\delta\beta=\nonumber \\&&
-\left({\partial\rho\over\partial\beta}\right)_p 
\left[{1\over\rho}Q^{ab}\nabla_ap \nabla_b
\delta U+(\omega+m\Omega)\bigl(\delta U + \delta \Phi\bigr)\right],
\nonumber \\\label{3.6}
\eeqa

\beqa
\nabla^a\nabla_a&&\delta\Phi +4\pi G \rho
\left({\partial \rho\over \partial p}\right)_\beta\delta \Phi= \nonumber \\
&&-4\pi G\rho\left({\partial\rho\over\partial p}\right)_\beta\delta U 
-4\pi G\left({\partial\rho\over\partial
\beta}\right)_p\delta\beta.\label{3.7}
\eeqa

The functions $\delta U$, $\delta\beta$, and $\delta\Phi$ are also
subject to appropriate boundary conditions at the interface between
the superfluid core and the ordinary-fluid envelope, at the surface of
the star, and at infinity.  First, we consider the boundary at the
interface between the superfluid core and the ordinary-fluid envelope
of the star.  Mass and momentum conservation across this boundary
place a number of constraints on the continuity of these
functions~\cite{mendell-lind}.  In particular these conditions require
that the functions $\delta U$ and $\delta \Phi$ be continuous there.
In addition, $\nabla_a\delta\Phi$ must be continuous, while
$\nabla_a\delta U$ must have a discontinuity that is prescribed by

\beq
n^a[\nabla_a\delta U]_s - {1\over \rho^2}
\left({\partial\rho\over\partial\beta} \right)_p n^a\nabla_ap [\delta
\beta]_s
= n^a[\nabla_a\delta U]_o.\label{3.8}
\eeq

\noindent The subscripts $s$ and $o$ in Eq.~(\ref{3.8}) denote that
the quantities are to be evaluated as limits from the superfluid or
the ordinary-fluid side of the boundary respectively, and $n^a$
denotes the outward directed unit normal to the boundary surface.  The
function $\delta \beta$, which is of interest to us only within the
superfluid core, must satisfy the condition

\beqa
n^a[\nabla_a\delta\beta]_s &&-{4\gamma^2\Omega^2z^bn_b
\over (\omega + m\Omega)^2}
z^a[\nabla_a\delta\beta]_s\nonumber \\
&&\qquad\qquad+ {2m\gamma\Omega\varpi^b n_b\over (\omega+m\Omega)
\varpi}[\delta\beta]_s = 0,\label{3.9}
\eeqa

\noindent on the boundary of the superfluid core.  Here
we use the notation $\varpi$ for the cylindrical radial coordinate,
and $\varpi^a$ to denote the unit vector in the $\varpi$ direction.

Next consider the boundary conditions on the outer surface of the
star.  The function $\delta U$ must be constrained at this surface in
such a way that the Lagrangian perturbation in the pressure vanishes
there: $\Delta p =0$.  This condition can be written in terms of the
variables used here by noting that

\beq
\Delta p = \delta p +{\delta v^a\nabla_ap\over i(\omega+m\Omega)}, 
\label{3.10}
\eeq

\noindent where $\delta v^a$ is given in this region by
Eq.~(\ref{3.1}) with $\delta\beta=0$.  Thus using Eqs.~(\ref{2.7}) and
(\ref{3.1}) this boundary condition can be written in terms of $\delta
U$ and $\delta \Phi$ as

\beq
0=\biggl[\rho\,(\omega+m\Omega)(\delta U +\delta\Phi)
+ Q^{ab}\nabla_a p \nabla_b\delta U
\biggr]_o.\label{3.11}
\eeq

Finally, the perturbed gravitational potential $\delta\Phi$ must fall
off at infinity faster than $1/r$ in order that the mass of the
perturbed star remain the same as that of the equilibrium star:
$\lim_{\,r\rightarrow \infty}(r\delta\Phi) = 0$.  In addition
$\delta\Phi$ and its first derivative must be continuous at the
surface of the star.  The problem of finding the modes of
``uniformly'' rotating superfluid stars is reduced therefore to
finding the solutions to Eqs.~(\ref{3.5}), (\ref{3.6}) and (\ref{3.7})
subject to the appropriate boundary conditions including in particular
Eqs.~(\ref{3.8}), (\ref{3.9}), and (\ref{3.11}).

The equations for the potentials $\delta U$ and $\delta\beta$,
Eqs.~(\ref{3.5}) and (\ref{3.6}), have complicated dependences on the
frequency of the mode and the angular velocity of the star through
$Q^{ab}$ and $\tilde{Q}^{ab}$, as given in Eqs.~(\ref{3.3}) and
(\ref{3.4}).  In the analysis that follows it will be necessary to
have those dependences displayed more explicitly.  Here we are
interested in investigating the superfluid versions of $r-$modes.
Such modes have frequencies that go to zero linearly as the angular
velocity of the star vanishes.  Thus, it will be useful to define the
dimensionless frequency parameter $\kappa$:

\beq
\kappa\Omega = \omega + m\Omega.\label{3.12}
\eeq

\noindent The parameter $\kappa$ remains finite in the zero
angular-velocity limit for these modes.  Using this parameter and the
expressions for $Q^{ab}$ and $\tilde{Q}^{ab}$ from Eqs.~(\ref{3.3})
and (\ref{3.4}), we re-write Eqs.~(\ref{3.5}) and (\ref{3.6}) to
obtain the following equivalent forms:

\beqa
\nabla_a &&\Bigl[\rho (\kappa^2\delta ^{ab}- 4z^az^b)\nabla_b\delta U\Bigr]
+ {2m\kappa\over \varpi}\varpi^a\nabla_a\rho\,\, \delta U =\nonumber\\&&
-\kappa^2(\kappa^2-4)\Omega^2 \left[\rho\left({\partial\rho\over \partial p}
\right)_\beta (\delta U + \delta \Phi)+
\left({\partial\rho\over\partial\beta}
\right)_p\delta\beta\right]\nonumber\\
&&+\nabla_a\left[{1\over\rho}\left({\partial\rho\over\partial\beta}\right)_p
\delta\beta\, (\kappa^2\delta^{ab}-4z^az^b)\nabla_bp\right]\nonumber\\
&&-{2m\kappa\over\rho\varpi}\left({\partial\rho\over\partial\beta}
\right)_p\delta\beta\, \varpi^a\nabla_ap,
\label{3.13}
\eeqa
\vfill\break
\beqa
&&\nabla_a\left[{\tilde{\rho}\over \kappa^2-4\gamma^2}
(\kappa^2\delta^{ab}-4\gamma^2 z^az^b)
\nabla_b\delta\beta\right] \nonumber \\
&&\,\,\,
+\left[{2m\kappa\over\varpi} \varpi^a\nabla_a\left({\gamma\tilde{\rho}\over
\kappa^2-4\gamma^2}\right)
+\kappa^2\Omega^2 {\rho_n^2\over \rho}{\partial\over\partial\beta}
\left({\rho_p\over\rho_n}\right)_p\right]\delta\beta\nonumber\\
&&\,\,\,-{1\over \rho^3(\kappa^2-4)} 
\left({\partial\rho\over\partial\beta}\right)^2_p
(\kappa^2\delta^{ab}-4z^az^b)\nabla_ap\nabla_bp\,\delta\beta \nonumber \\
&&\quad=-\left({\partial\rho\over\partial\beta}\right)_p
\biggl[{\kappa^2\delta^{ab}-4z^az^b\over \rho(\kappa^2-4)}
\nabla_ap\nabla_b\delta U\nonumber\\
&&\qquad\qquad+{2m\kappa\over \rho(\kappa^2-4)\varpi}
\varpi^a\nabla_ap\, \delta U+\kappa^2\Omega^2(\delta U +\delta \Phi)
\biggr]\nonumber \\\label{3.14}
\eeqa

\noindent The boundary conditions, Eqs.~(\ref{3.9}) and
(\ref{3.11}) are similarly transformed into the forms:

\beqa
\kappa^2n^a[\nabla_a\delta\beta]_s -4\gamma^2z^bn_bz^a
[\nabla_a\delta\beta]_s
&&+ {2m\kappa\gamma\varpi^b n_b\over
\varpi}[\delta\beta]_s = 0.\nonumber \\\label{3.15}
\eeqa

\beqa
\biggl[\bigl(\kappa^2\delta^{ab} &&- 4z^az^b\bigr)\nabla_ah\nabla_b\delta U
+{2m\kappa\over\varpi} \varpi^a\nabla_ah\,\,\delta U\nonumber\\
&&+\kappa^2(\kappa^2-4)\Omega^2(\delta U +\delta\Phi)
\biggr]_{o}=0.\label{3.16}
\eeqa

\noindent In Eq.~(\ref{3.16}) we have expressed the boundary condition
in terms of the thermodynamic enthalpy, $h$, which is defined as

\beq
h(p)=\int_0^p{dp'\over \rho(p')}.\label{3.17}
\eeq

\noindent The enthalpy is the appropriate thermodynamic function to
use in Eq.~(\ref{3.16}) because its gradient, $\nabla_a h$, is
well-defined and non-zero at the surface of the star.

%
\section{Superfluid $r-$Modes in the Slow
Rotation Approximation}
\label{sectionIV}

The $r$-modes of rotating ordinary-fluid stars have traditionally been
studied using a small angular-velocity expansion~\cite{p&p}.  Our goal
here is to perform a similar expansion for the superfluid
generalizations of the $r$-modes.  Thus, we seek solutions to
Eqs. (\ref{3.13}), (\ref{3.14}) and (\ref{3.7}) as power series in the
angular velocity of the star.  

We begin first with the structure of the equilibrium superfluid star.
This structure is identical to its ordinary-fluid counterpart in the
large-scale averaged-over-vortices hydrodynamics used here.  We expand
each of the equilibrium functions of interest:

\beq
\rho=\rho_0+\rho_2{\Omega^2\over \pi G \bar{\rho}_0}
+ {\cal O}(\Omega^4),
\label{4.1}
\eeq
\beq
p=p_0+p_2{\Omega^2\over \pi G \bar{\rho}_0}
+ {\cal O}(\Omega^4),
\label{4.2}
\eeq
\beq
h=h_0+h_2{\Omega^2\over \pi G \bar{\rho}_0}
+ {\cal O}(\Omega^4).
\label{4.3}
\eeq

\noindent The location of the surface of the star $R(\mu)$
is also expressed as such an expansion:

\beq
R=R_0+R_2{\Omega^2\over \pi G \bar{\rho}_0}
+ {\cal O}(\Omega^4).
\label{4.4}
\eeq

\noindent Here and throughout the remainder of this paper we use the
subscripts $0$ and $2$ to denote the lowest- and the second-order
terms respectively in these expansions; and we use $r$ and
$\mu=\cos\theta$ to denote the standard spherical coordinates.  We
also introduce here the angular velocity scale $\sqrt{\pi G
\bar{\rho}_0}$, where $\bar{\rho}_0$ is the average density of the
star in the non-rotating limit.  (Equilibrium neutron-star models do
not exist for $\Omega\gtrsim \case{2}{3}\sqrt{\pi G \bar{\rho}_0}$.)
The techniques needed to evaluate the terms in these series for the
equilibrium structure are identical to those described for example in
Lindblom, Mendell, and Owen~\cite{lmo}.

Next, we define expansions for the quantities that determine the
perturbations of a superfluid star, $\delta U$, $\delta \beta$,
$\delta\Phi$ and $\kappa$:

\beq \delta U = R_0^2\Omega^2\left[\delta U_0 + \delta U_2 
{\Omega^2\over \pi G \bar{\rho}_0} + {\cal O}(\Omega^4)\right],\label{4.5} 
\eeq

\beq \delta \beta 
= R_0^2\Omega^2\left[\delta \beta_0 + \delta \beta_2 {\Omega^2\over \pi G
\bar{\rho}_0} + {\cal O}(\Omega^4)\right],\label{4.6} \eeq

\beq
\delta \Phi = R_0^2\Omega^2\left[\delta \Phi_0 
+ \delta \Phi_2 {\Omega^2\over \pi G \bar{\rho}_0}
+ {\cal O}(\Omega^4)\right],\label{4.7}
\eeq

\beq
\kappa=\kappa_0 + \kappa_2 
{\Omega^2\over \pi G \bar{\rho}_0}
+ {\cal O}(\Omega^4).\label{4.8}
\eeq

\noindent We have normalized the eigenfunctions using $R_0$, the
radius of the star (in the non-rotating limit), and $\Omega$, the
angular velocity of the star.  Using these expressions for the
perturbations, together with those for the structure of the
equilibrium star, it is straightforward to write down order-by-order
the equations that determine the superfluid $r$-modes.  This expansion
is completely analogous to that given by Lindblom, Mendell, and
Owen~\cite{lmo} for the ordinary-fluid modes.

It is straightforward to verify that the functions

\beq
\delta U_0 = \alpha \left ( {r \over R_0} \right )^{m+1} 
P^m_{m+1}(\mu)e^{im\varphi},
\label{4.9}
\eeq

\beq
\delta\beta_0=0,\label{4.10}
\eeq

\noindent with
\beq
\kappa_0 = {2\over m+1},\label{4.11}
\eeq

\noindent 
satisfy the lowest-order terms from the expansion of the pulsation
Eqs.~(\ref{3.13}) and (\ref{3.14}).  Inspection of the equation for $\delta
\beta$, Eq.~(\ref{3.14}), reveals that the right side is proportional
to $\Omega^3$: the first term on the right vanishes identically for
$\delta U_0$ given by Eq.~(\ref{4.9}), while the lowest-order
contribution to the second term is proportional to $\Omega^3$.  Thus
the function $\delta\beta_0=0$ satisfies Eq.~(\ref{3.14}) to lowest
order.

With the lowest-order contribution to $\delta\beta$ vanishing, the
lowest-order equations for $\delta U$ and $\delta\Phi$ from
Eqs.~(\ref{3.13}) and (\ref{3.7}) reduce to the following,

\beq \nabla_a\Bigl[\rho_0 (\kappa_0^2\delta ^{ab} -
4z^az^b)\nabla_b\delta U_0\Bigr] + {2m\kappa_0\over
\varpi}\varpi^a\nabla_a\rho_0\,\, \delta U_0 =0, \label{4.12} \eeq

\beq
\nabla^a\nabla_a\delta\Phi_0 = - 4\pi G \left({d\rho\over dh}\right)_0
(\delta U_0 + \delta \Phi_0).
\label{4.13}
\eeq

\noindent The lowest-order boundary conditions at the
superfluid ordinary-fluid interface, Eqs.~(\ref{3.8}) and (\ref{3.9}),
merely require that $\delta U_0$ and $\nabla_a\delta U_0$ are
continuous there.  The lowest-order contribution from the boundary
condition for $\delta U_0$ at the surface of the star is

\beqa
\biggl[
\bigl(\kappa_0^2\delta^{ab} 
- &&4z^az^b\bigr)\nabla_ah_0\nabla_b\delta U_0\nonumber \\
&&+{2m\kappa_0\over\varpi} \varpi^a\nabla_ah_0\,\,\delta U_0
\biggr]_{r=R_0}=0.
\label{4.14}
\eeqa

\noindent These equations are identical the lowest-order terms in the
ordinary-fluid $r$-mode equations~\cite{lmo}.  The function $\delta
U_0$ together with $\kappa_0$ given in Eqs.~(\ref{4.9}) and
(\ref{4.11}) satisfy Eqs.~(\ref{4.12}) and (\ref{4.14}) identically
because these are in fact the lowest-order expressions for the
classical $r-$modes as studied for example by Papaloizou and
Pringle~\cite{p&p}. (However, they are expressed here in a form that
was introduced more recently~\cite{li98}.)  Thus to lowest order the
superfluid $r$-modes are identical to their ordinary-fluid
counterparts.

Continuing on to second order, the equations for the potentials are

\beqa
\nabla_a\Bigl[&&\rho_0(\kappa_0^2\delta^{ab}-4z^az^b)\nabla_b\delta
U_2\Bigr] +{2m\kappa_0\over \varpi}\varpi^a\nabla_a\rho_0\,\,\delta
U_2\nonumber\\
&&+\nabla_a\Bigl[\rho_2(\kappa_0^2\delta^{ab}-4z^az^b)\nabla_b\delta
U_0 +2\kappa_0\kappa_2\rho_0\nabla^a\delta U_0\Bigr]\nonumber \\
&&+{2m\over\varpi}\varpi^a\left(\kappa_2\nabla_a\rho_0
+\kappa_0\nabla_a\rho_2\right)\delta U_0 =\nonumber \\ &&\qquad -\pi G
\bar{\rho}_0\kappa_0^2(\kappa_0^2-4) \rho_0\left({\partial\rho\over
\partial p}\right)_\beta (\delta U_0 +\delta\Phi_0)\nonumber\\
&&\qquad + \nabla_a\left[{1\over \rho_0}
\left({\partial\rho\over\partial\beta}\right)_p\delta\beta_2\,
(\kappa_0^2\delta^{ab}-4z^az^b)\nabla_b p_0\right]\nonumber \\ &&\qquad
-{2m\kappa_0\over\varpi}{1\over\rho_0}
\left({\partial\rho\over\partial\beta}\right)_p \delta\beta_2
\,\varpi^a\nabla_a p_0,
\label{4.15}
\eeqa

\beqa
&&\nabla_a\left[{\tilde{\rho}_0\over \kappa^2_0-4\gamma^2_0}
(\kappa^2_0\delta^{ab}-4\gamma^2_0 z^az^b)
\nabla_b\delta\beta_2\right] \nonumber \\
&&\quad
+{2m\kappa_0\over\varpi} \varpi^a\nabla_a\left({\gamma_0\tilde{\rho}_0
\over\kappa^2_0-4\gamma^2_0}\right)\delta\beta_2\nonumber \\
&&\quad-{1\over \rho^3_0(\kappa^2_0-4)} 
\left({\partial\rho\over\partial\beta}\right)^2_p
(\kappa^2_0\delta^{ab}-4z^az^b)\nabla_ap_0\nabla_bp_0\,\delta\beta_2 
\nonumber \\
&&\quad=-{1\over  \rho_0(\kappa^2_0-4)} 
\left({\partial\rho\over\partial\beta}\right)_p
\biggl[(\kappa^2_0\delta^{ab}-4z^az^b)
\nabla_ap_0\nabla_b\delta U_2\nonumber\\
&&\qquad\quad+{2m\kappa_0\over \varpi}
\varpi^a\nabla_ap_0\, \delta U_2+2\kappa_0\kappa_2
\nabla^ap_0\nabla_a\delta U_0\nonumber \\
&&\qquad\quad
+{2m\kappa_2\over \varpi}\varpi^a\nabla_ap_0\, \delta U_0
+{2m\kappa_0\over \varpi}\varpi^a\nabla_ap_2\, \delta U_0\nonumber \\
&&\qquad\quad+(\kappa^2_0\delta^{ab}-4z^az^b)
\nabla_ap_2\nabla_b\delta U_0\biggr]\nonumber\\
&&\qquad\quad
-\kappa^2_0\pi G \bar{\rho}_0
\left({\partial\rho\over\partial\beta}\right)_p
(\delta U_0 +\delta \Phi_0),\label{4.16}
\eeqa

\beqa \nabla^a&&\nabla_a\delta\Phi_2 +4\pi G \rho_0
\left({\partial\rho\over \partial
p}\right)_\beta\delta\Phi_2=\nonumber\\ &&-4\pi G \rho_0
\left({\partial\rho\over \partial p}\right)_\beta \delta U_2 -4\pi G
\left({\partial\rho\over \partial\beta}\right)_p\delta\beta_2
\nonumber \\
&&-4\pi G\left({\partial\rho\over \partial
p}\right)_\beta^{-1}{\partial\over \partial p}\left[
\rho_0\left({\partial\rho\over \partial p}\right)_\beta
\right]_\beta\rho_2
(\delta U_0 + \delta \Phi_0).\nonumber \\\label{4.17} \eeqa

It will be helpful to express Eqs.~(\ref{4.15}) and (\ref{4.16}) in
the following shorthand forms:

\beq
D(\delta U_2)=E(\delta \beta_2) + F,\label{4.18}
\eeq

\beq
\tilde{D}(\delta \beta_2) = \tilde{E}(\delta U_2) + \tilde{F},
\label{4.19}
\eeq

\noindent where $D$ and $\tilde{D}$ are second-order partial
differential operators, $E$ and $\tilde{E}$ are first-order operators,
and $F$ and $\tilde{F}$ are functions that depend on the lowest-order
eigenfunctions $\delta U_0$ and $\delta\Phi_0$.  

We next consider the second-order boundary conditions.  The boundary
conditions that must be satisfied at the superfluid ordinary-fluid
boundary, Eqs.~(\ref{3.8}) and (\ref{3.9}), have the following
second-order forms on the surface $r=R_s$:

\beq
\left[{\partial\delta U_2\over\partial r}\right]_s
-{1\over \rho_0^2}\left({\partial\rho\over\partial\beta}\right)_p
{dp_0\over dr}\left[\delta\beta_2\right]_s =
\left[{\partial\delta U_2\over\partial r}\right]_o,
\label{4.20}
\eeq

\beqa
(\kappa_0^2 - 4\gamma_0^2\mu^2)
&&\left[{\partial\delta\beta_2\over\partial r}\right]_s
- {4\gamma_0^2\over R_s} \mu (1 - \mu^2)
\left[{\partial\delta\beta_2\over\partial \mu}\right]_s
\nonumber \\
&&+{2m\kappa_0\gamma_0\over R_s}\left[\delta\beta_2\right]_s=0.
\label{4.21}
\eeqa

\noindent The second-order boundary condition for the potential $\delta U$,
Eq.~(\ref{3.16}), is identical to that derived by Lindblom,
Mendell, and Owen~\cite{lmo} for the perfect-fluid case:

\beqa
\Biggl\{(&&\kappa_0^2\delta^{ab} - 4z^az^b)\nabla_ah_0\nabla_b\delta U_2 +
{2m\kappa_0\over \varpi}\varpi^a\nabla_ah_0\delta U_2\nonumber \\
&&+(\kappa_0^2\delta^{ab} - 4z^az^b)\nabla_ah_2\nabla_b\delta U_0 +
{2m\kappa_0\over \varpi}\varpi^a\nabla_ah_2\delta U_0\nonumber \\
&&+2\kappa_0\kappa_2\nabla^ah_0\nabla_a\delta U_0
+{2m\kappa_2\over \varpi}\varpi^a\nabla_a h_0\delta U_0\nonumber \\
&&+\kappa_0^2(\kappa_0^2-4)\pi G\bar{\rho}_0(\delta U_0+\delta\Phi_0)
\Biggr\}_{r=R_0}=0.
\label{4.22}
\eeqa

It will be useful in our numerical solution of the pulsation equations
to have an expression for the second-order contribution to the
frequency of the mode, $\kappa_2$, as integrals over the
eigenfunctions.  Such an expression can be obtained by multiplying
Eq.~(\ref{4.18}) by $\delta U_0^*$ and integrating over the interior
of the star.  Since $\delta U_0^*$ is an element of the kernel of the
operator $D$, this part of the integral reduces to a boundary
integral.  This boundary integral is non-vanishing because
$\nabla_a\delta U_2$ is discontinuous at the superfluid boundary as a
consequence of the boundary condition Eq.~(\ref{4.20}).  The result of
integrating the left side of Eq.~(\ref{4.18}) can be expressed
therefore in the following way:

\beqa
&&\int\delta U_0^* D(\delta U_2) d^{\,3}x =\nonumber\\
&&
\qquad2\pi \biggl[r^2 \left({\partial\rho\over\partial\beta}\right)_p
{dh_0\over dr} \int^1_{-1}(\kappa_0^2-4\mu^2)\delta\beta_2\delta U_0^*
d\mu\biggr]_{r=R_s}.\nonumber\\
\label{4.23}
\eeqa

\noindent Combining this result with the more straightforward integrals
of the right side of Eq.~(\ref{4.18}), we obtain the following expression
for the second-order change in the frequency of the superfluid $r$-mode:

\beqa
&&\kappa_2\int {1\over r} {d\rho_0\over dr}
|\delta U_0|^2 d^{\,3}x= {6m\over (m+1)^2}\int
{\rho_{22}\over r^2}|\delta U_0|^2d^{\,3}x\nonumber\\
&&\quad+{8\pi G\bar{\rho}_0 m\over (m+1)^4}\int
\left({d\rho\over dh}\right)_0
(\delta U_0+\delta\Phi_0)\delta U_0^* d^{\,3}x\nonumber\\
&&\quad+\int{(\kappa_0^2-4\mu^2)r\over 2(m+2)}
{d\over dr}\left[{1\over r}{dh_0\over
dr}\left({\partial\rho\over\partial\beta}
\right)_p\right]\delta \beta_2\delta U_0^* d^{\,3}x\nonumber\\
&&\quad+\int{dh_0\over dr}\left({\partial\rho\over\partial\beta}
\right)_p\Biggl[{\kappa_0^2-4\mu^2\over 2(m+2)}
{\partial\delta\beta_2\over \partial r}\nonumber\\
&&\qquad\qquad\qquad\qquad\qquad
-{2\mu(1-\mu^2)\over (m+2)r}{\partial\delta\beta_2\over\partial\mu}\Biggr]
\delta U_0^* d^{\,3}x\nonumber\\
&&\quad+{3\kappa_0^2-2m\kappa_0-4\over2(m+2)}\int{1\over r}{dh_0\over dr}
\left({\partial\rho\over\partial\beta}\right)_p\delta \beta_2\delta U_0^*
d^{\,3}x\nonumber\\
&&\quad-\left[\pi r^2{dh_0\over dr}
\left({\partial\rho\over\partial\beta}\right)_p
\int{\kappa_0^2-4\mu^2\over m+2}
\delta\beta_2\delta U_0^* d\mu\right]_{r=R_s}\!\!\!\!\!.
\label{4.24}
\eeqa

\noindent In Eq.~(\ref{4.24}) the integrations are to be performed
over the interior of the superfluid core, $0\leq r\leq R_s$, for those
integrals involving $\delta \beta_2$ and throughout the interior of
the star, $0\leq r\leq R_0$, for those integrals that do not.  Even
though this expression for $\kappa_2$ depends on the second-order
eigenfunction $\delta\beta_2$ it is nevertheless very useful in
determining $\kappa_2$ numerically. This is due to the fact that the
integrals involving  $\delta\beta_2$ are typically quite small.
\section{Numerical Solutions for the Superfluid $r$-Modes}
\label{sectionV}

In order to solve the equations for the superfluid $r$-modes we must
adopt a specific model for the equilibrium structure of the neutron
star, and explicit expressions for the various thermodynamic functions
that appear in the equations.  For the equilibrium structure of the
neutron star, we use the simple polytropic equation of state:
$p=K\rho^2$, with $K$ chosen so that a $1.4M_\odot$ model has a radius
of $R_0=12.533$ km.  We choose this simple model because our method of
solving Eq.~(\ref{4.18}) to determine $\delta U_2$ seems to be rather
unstable for more realistic equations of state.  We also use this
simple equilibrium equation of state to evaluate the thermodynamic
derivative $(\partial\rho/\partial p)_\beta =\rho/2p$ that appears in
the pulsation equations.  We adopt the fiducial value
$\rho_s=2.8\times 10^{14}$~g/cm${}^3$ for the superfluid transition
density.  However, we examine the sensitivity of our results to
variations in this parameter.

In order to evaluate the other thermodynamic properties of neutron
star matter needed in the pulsation equations, we use the recent
semi-empirical equation of state, A18$+\delta v+$UIX, of Akmal,
Pandharipande, and Ravenhall~\cite{apr}.  The derivative
$(\partial\rho/\partial \beta)_p$ is determined from this equation of
state using the formula, Eq.~(70), in Lindblom and
Mendell~\cite{mendell-lind} that expresses this derivative in terms of
the more easily evaluated derivatives of the chemical potentials. For
the density range of primary interest to us ($2.8\times
10^{14}\leq\rho\leq 10^{15}$ gm/cm${}^3$), we find that

\beq
\left({\partial\rho\over\partial\beta}\right)_p \approx
1.4\times10^{-7} - 1.1\times10^{-22}\rho,
\label{5.1}
\eeq

\noindent where all quantities are expressed in cgs units.  Similarly,
we find that the proton fraction $\rho_p/\rho$ for this equation of
state is given approximately by

\beq
{\rho_p\over\rho}\approx 0.031 + 8.8\times 10^{-17}\rho.
\label{5.2}
\eeq

\noindent These expressions are accurate to within about $20\%$ in
the indicated density range.

The thermodynamic quantities $\tilde{\rho}$ and $\gamma$ can be
expressed in terms of the baryon densities $\rho_n$ and $\rho_p$, and
the mass-density matrix element $\rho_{np}$ quite
generally using the constraints on the mass-density matrix 
from Galilian invariance and Eq.~(\ref{2.1}):

\beq
\gamma=1+\epsilon{\rho\over\rho_p},\label{5.5}
\eeq

\beq
\tilde{\rho}=\rho\left(1-{\rho_p\over\rho}\right)
\left(\epsilon+{\rho_p\over\rho}\right).\label{5.6}
\eeq

\noindent Thus using Eq.~(\ref{5.2}) for the proton fraction, we
obtain simple smooth expressions for the superfluid thermodynamic
functions needed in Eqs.~(\ref{4.18}) and (\ref{4.19}).

In order to evaluate the potentials $\delta U_2$ and $\delta\beta_2$,
we convert the operators in Eqs.~(\ref{4.18}) and (\ref{4.19}) into
matrix operators by making the usual discrete representations of the
various derivatives that appear.  To this end we have given
expressions for these operators in spherical coordinates in the
Appendix.  We convert the radial derivatives into discrete form using
the standard three-point differencing formulae.  For the angular
derivatives, we have written separate codes that use either the
higher-order angular differencing formulae discussed in Ipser and
Lindblom~\cite{ipser-lind} or the standard three-point formulae.  We
find that the results of our two separate codes agree very well.  The
code that uses the higher-order angular differencing formulae requires
far fewer angular spokes in order to achieve a given accuracy, and it
appears to be somewhat more stable.

The operator $D$ that appears in Eq.~(\ref{4.18}) is a hyperbolic
differential operator with a non-trivial kernel.  Thus,
Eq.~(\ref{4.18}) is solved using the numerical relaxation technique
developed in Lindblom, Mendell, and Owen~\cite{lmo}.  The operator
$\tilde{D}$ is also hyperbolic, but (generically) it does not have a
non-trivial kernel.  We find that $\tilde{D}$ can be inverted
numerically without difficulty.  Thus, Eq.~(\ref{4.19}) can be solved
by straightforward numerical techniques.

The Eqs.~(\ref{4.18}) and (\ref{4.19}) are solved iteratively to
determine the potentials $\delta U_2$ and $\delta \beta_2$.  We begin
the process by setting $\delta \beta_2=0$.  Then we iterate the
following sequence of calculations. First, we evaluate $\kappa_2$ from
Eq.~(\ref{4.24}); then we solve Eq.~(\ref{4.18}) for $\delta U_2$; and
finally we solve Eq.~(\ref{4.19}) for $\delta \beta_2$.  This sequence
of calculations is iterated until the functions $\delta U_2$ and
$\delta \beta_2$ converge.  This generally takes only a few steps.  We
find that the second-order eigenvalue $\kappa_2$ has the value
$0.29879$ for this model, essentially independent of the entrainment
parameter $\epsilon$ and the superfluid-transition density $\rho_s$.
This value of $\kappa_2$ differs only slightly from the value $0.29883$
found by Lindblom, Mendell, and Owen~\cite{lmo} for the analogous
ordinary-fluid $r$-modes.

\bfig \centerline{\psfig{file=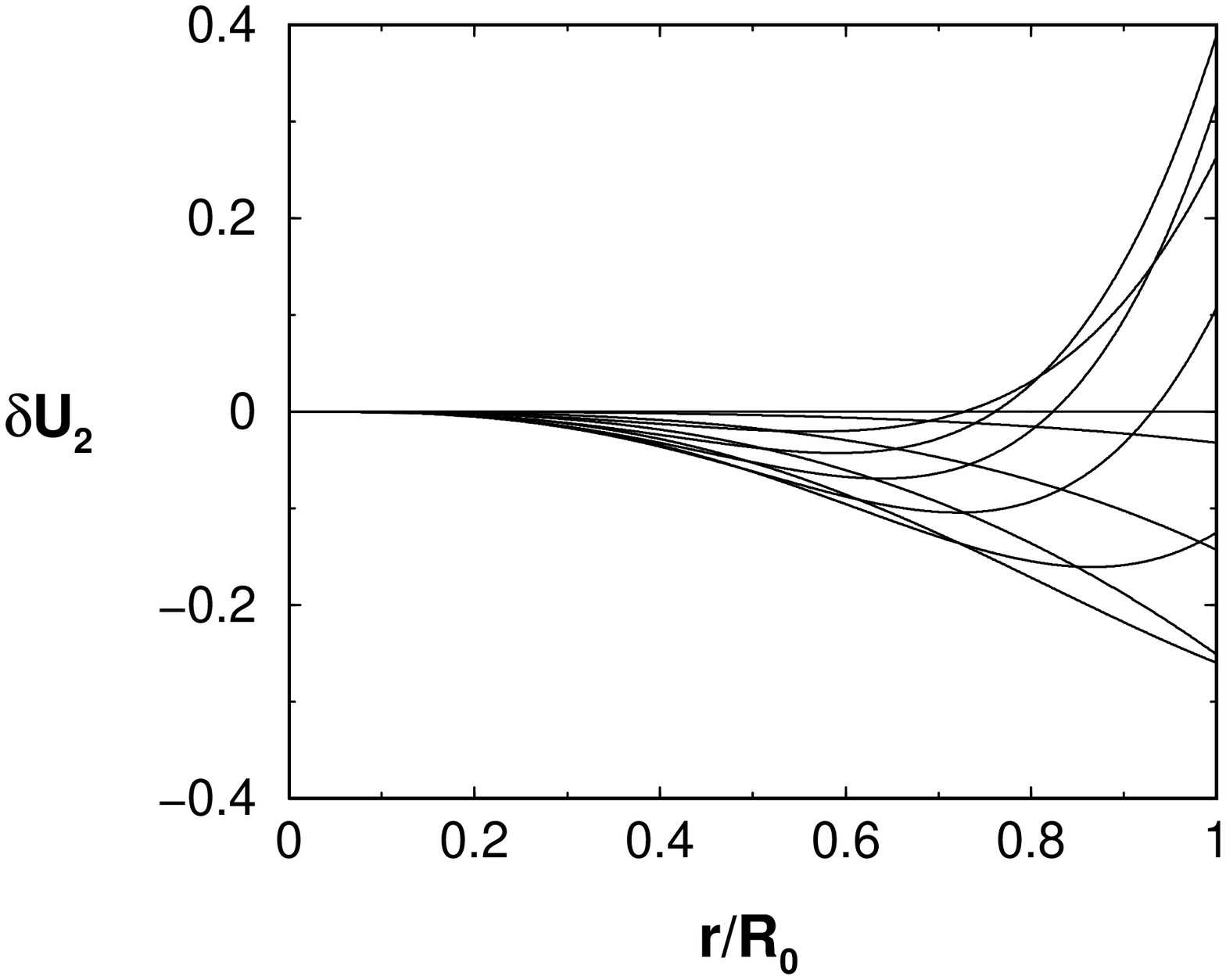,height=2.4in}} \vskip 0.3cm
\caption{Eigenfunction $\delta U_2$ for the $m=2$ $r$-mode
based on a superfluid model with $\epsilon=0.04$ and 
$\rho_s=2.8\times10^{14}$ g$/$cm${}^3$.\label{fig1}} \efig

The eigenfunction $\delta U_2$ for the case $\epsilon=0.04$ and
$\rho_s=2.8\times 10^{14}$g/cm${}^3$ is illustrated in
Fig.~\ref{fig1}.  This function was computed on a grid having 2000
radial grid points and 10 angular spokes using the higher-order
angular differencing code.  Each curve represents the radial
dependence of $\delta U_2$ along one of the angular spokes.  This
function is essentially identical to that obtained by Lindblom,
Mendell, and Owen~\cite{lmo} for the ordinary-fluid case.
Fig.~\ref{fig2} illustrates the associated function $\delta\beta_2$
that was evaluated together with the $\delta U_2$ of Fig.~\ref{fig1}.
We note that $\delta\beta_2$ is only defined within the core of the
star where the neutron-star matter is superfluid.  We also note that
$\delta\beta_2$ is about an order of magnitude smaller than $\delta
U_2$.  Thus even at the second-order, the superfluid $r$-modes differ
little from their ordinary-fluid counterparts.

\bfig \centerline{\psfig{file=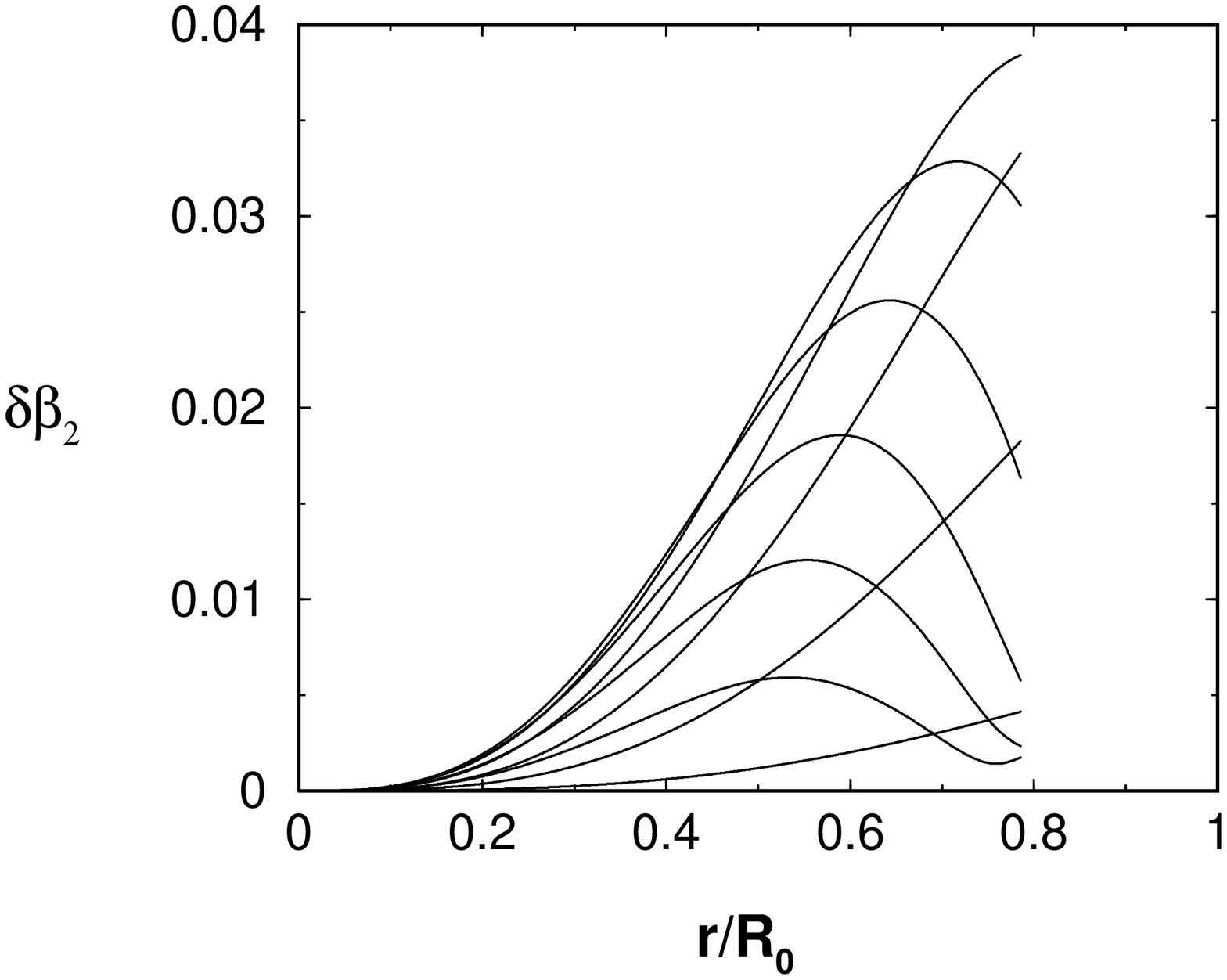,height=2.4in}} \vskip 0.3cm
\caption{Eigenfunction $\delta\beta_2$ for the $m=2$ $r$-mode
based on a superfluid model with $\epsilon=0.04$ and 
$\rho_s=2.8\times10^{14}$ g$/$cm${}^3$.\label{fig2}} \efig

\bfig \centerline{\psfig{file=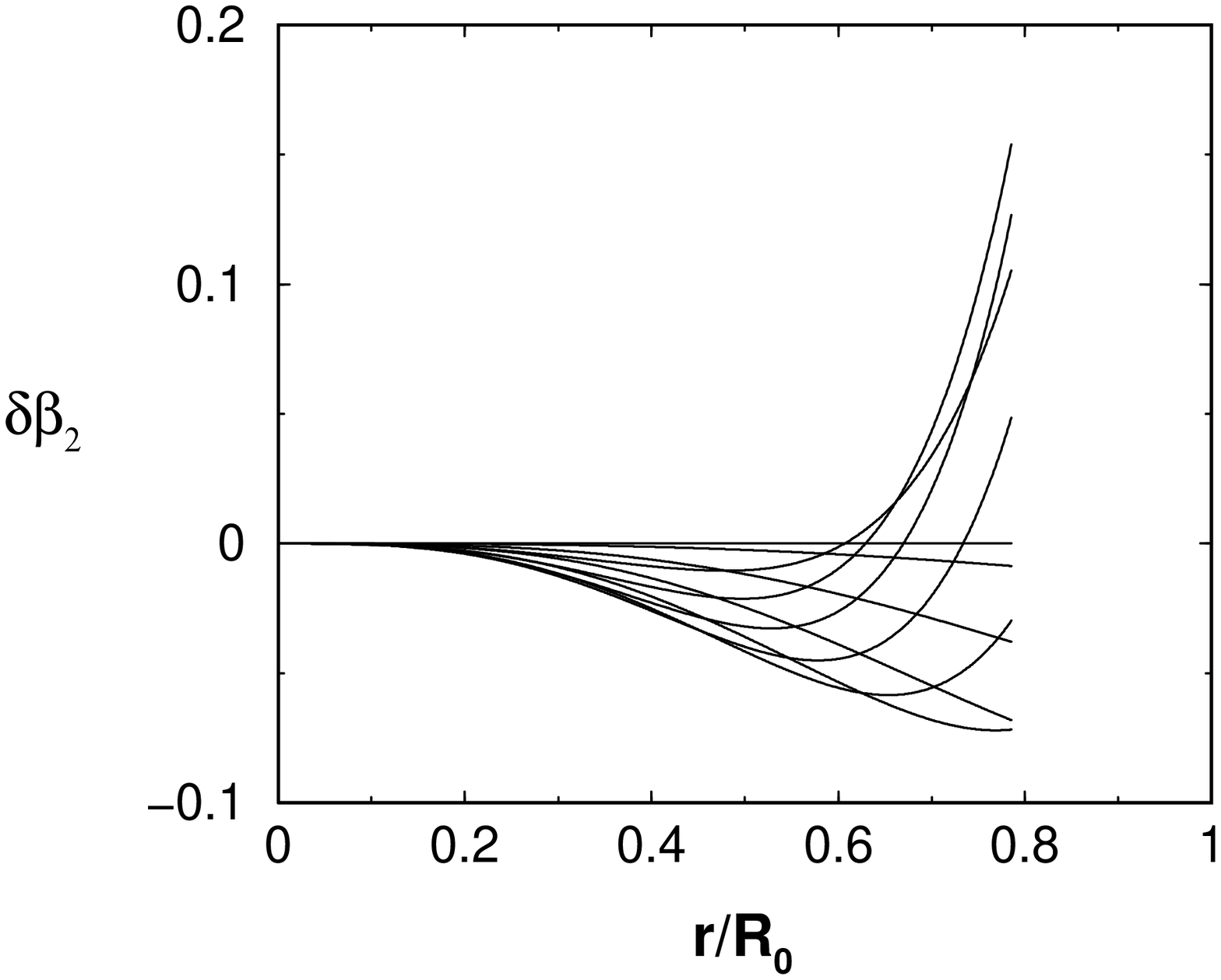,height=2.4in}} \vskip 0.3cm
\caption{Eigenfunction $\delta\beta_2$ for the $m=2$ $r$-mode
based on a superfluid model with $\epsilon=0.02$ and 
$\rho_s=2.8\times10^{14}$ g$/$cm${}^3$.\label{fig3}} \efig

Figures~\ref{fig3} and \ref{fig4} illustrate how $\delta\beta_2$
changes as $\epsilon$ varies.  In particular Figs.~\ref{fig3} and
\ref{fig4} illustrate $\delta\beta_2$ for the extreme values of
$\epsilon$ considered here: $\epsilon=0.02$ and $\epsilon=0.06$
respectively.  While the differences in these three functions,
Figs.~\ref{fig2}, \ref{fig3}, and \ref{fig4}, are significant, they do
not really illustrate the most interesting feature of their dependence
on $\epsilon$.  The most interesting and unexpected feature that we
find in the solutions for $\delta\beta_2$ is a kind of
resonance phenomenon.  We find that there are certain ``critical''
values of $\epsilon$ ($\epsilon_c=0.02294$ and $\epsilon_c=0.04817$ for
the $\rho_s=2.8\times 10^{14}$g/cm${}^3$ case) near which the function
$\delta\beta_2$ becomes extremely large.  Near these special values
of $\epsilon$ the character of the second-order terms in the expansion
for the mode change from being dominated by the ordinary-fluid like
correlated motion of the neutrons and protons, to the uniquely
superfluid anti-correlated motion, where the neutrons and protons have
opposite velocities.  We see a hint of this behavior,
perhaps, in Fig.~\ref{fig3} where the magnitude of $\delta\beta_2$ for
$\epsilon=0.02$ is considerably larger than its value at
$\epsilon=0.04$ or 0.06.

\bfig \centerline{\psfig{file=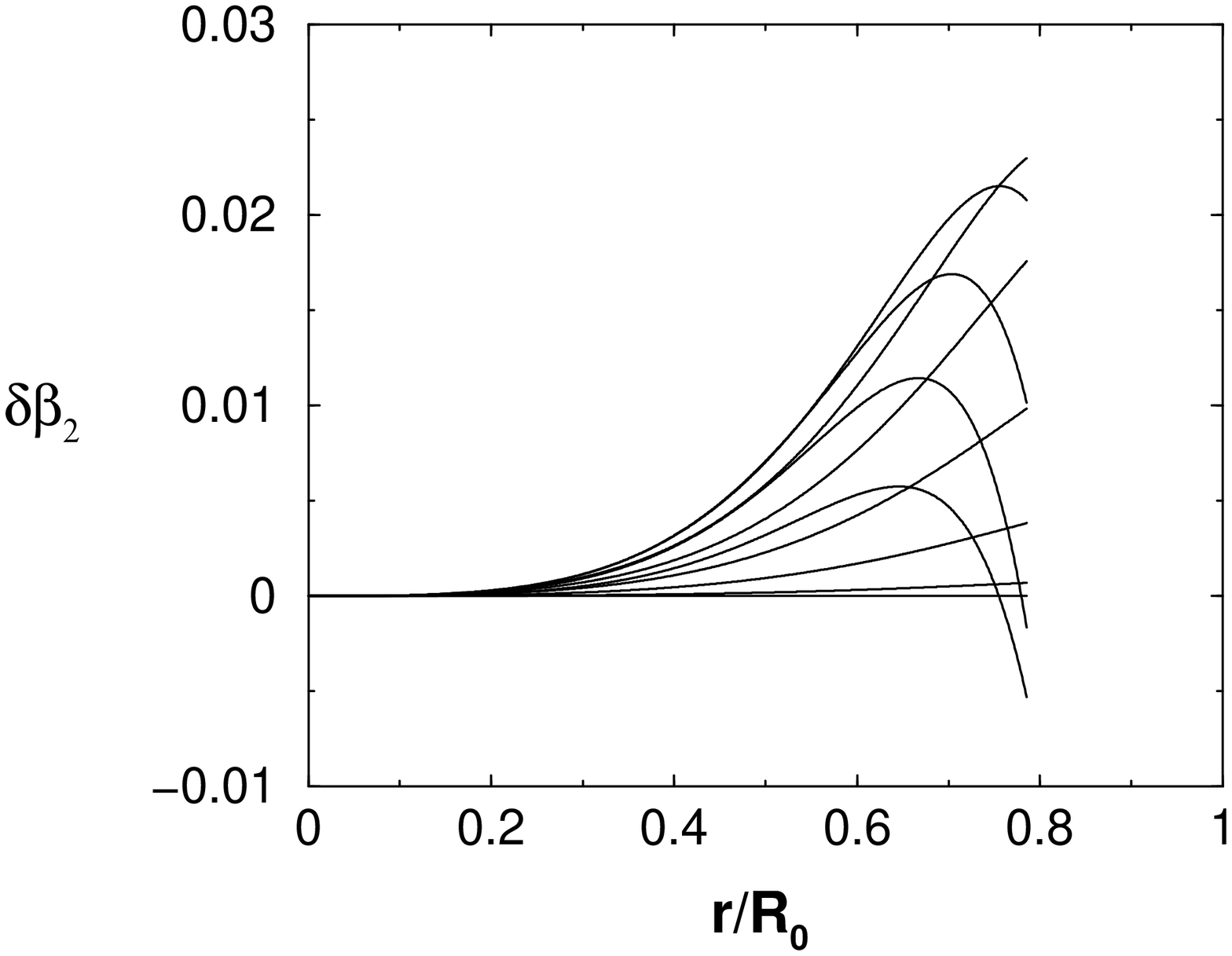,height=2.4in}} \vskip 0.3cm
\caption{Eigenfunction $\delta\beta_2$ for the $m=2$ $r$-mode
based on a superfluid model with $\epsilon=0.06$ and 
$\rho_s=2.8\times10^{14}$ g$/$cm${}^3$.\label{fig4}} \efig

Mathematically we find that the reason for this resonance phenomenon
is that the operator $\tilde{D}$ that determines $\delta\beta_2$ in
Eq.~(\ref{4.19}) becomes singular at these critical values of
$\epsilon$.  This singular behavior is illustrated in Fig.~\ref{fig5},
in which we depict the dependence of the smallest (in absolute value)
eigenvalue of the operator $\tilde{D}$ as a function of $\epsilon$.
We see that this smallest eigenvalue vanishes for the two critical
values of $\epsilon$ noted above: $\epsilon_c=0.02294$ and
$\epsilon_c=0.04817$.  In a sufficiently small neighborhood of these
points the function $\delta\beta_2$ must be proportional therefore to
$\psi_{\min}/\lambda_{\min}$ where $\psi_{\min}$ is the eigenfunction
of $\tilde{D}$ that corresponds to this smallest eigenvalue,
$\lambda_{\min}$.  Since $\lambda_{\min}$ vanishes for the singular
values of $\epsilon$, this forces the function $\delta\beta_2$ to
become very large in sufficiently small neighborhoods of these points.
While this vanishing of the eigenvalues of $\tilde{D}$ explains
mathematically why $\delta\beta_2$ becomes large for certain values of
$\epsilon$, it does not really explain physically the cause of this
resonance phenomenon.

\bfig \centerline{\psfig{file=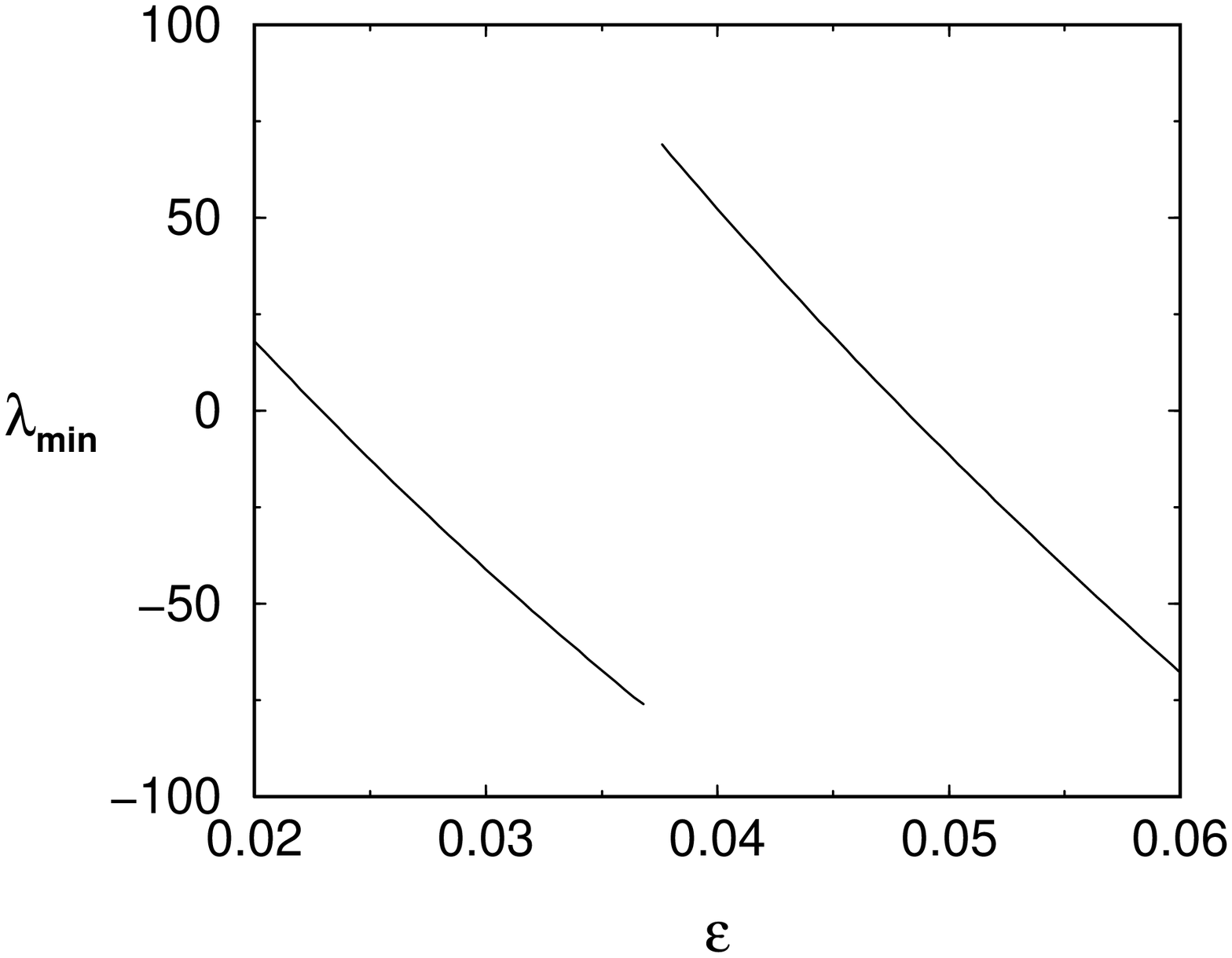,height=2.4in}} \vskip 0.3cm
\caption{Eigenvalue of the operator $\tilde{D}$ having the smallest
absolute value as a function of $\epsilon$, for $m=2$ perturbations
with $\rho_s=2.8\times10^{14}$ g$/$cm${}^3$.\label{fig5}} \efig

\section{Dissipation in  Superfluid $r$-Modes}
\label{sectionVI}

The primary motivation for our study here is to investigate how the
transition to a superfluid state at low temperatures effects the
stability of the $r$-modes in older colder neutron stars.  The
$r$-modes are driven towards instability by gravitational
radiation~\cite{andersson,fried-morsink}, but internal fluid
dissipation tends to suppress this instability~\cite{lom,aks}.  Internal
fluid dissipation can stabilize the $r$-modes completely, if it is
sufficiently strong.  In this section we investigate the importance of
the two types of internal fluid dissipation considered most likely to
have a substantial influence on the $r$-modes in superfluid neutron
stars: mutual friction caused by the scattering of electrons off the
cores of the neutron vortices, and shear viscosity due to
electron-electron scattering.

The effects of dissipation on the evolution of a fluid are most
conveniently studied using an appropriate energy functional.  For the
case of a neutron-star superfluid whose evolution is determined by
Eqs.~(\ref{2.5}), (\ref{2.6}), (\ref{2.11}), (\ref{2.12}), and
(\ref{2.13}), the following is the appropriate energy
functional~\cite{mendell1991b}

\beqa
{\cal E} = \case{1}{2}\int\biggl\{
&&\rho\,\delta v^*_a\delta v^a 
+\rho\left({\partial\rho\over\partial p}\right)_\beta
{\rm Re}[(\delta U+\delta\Phi)\delta U^*]\nonumber \\
&&+ \tilde{\rho}\,\delta w^*_a\delta w^a
+{\rho_n^2\over \rho}{\partial\over\partial\beta}
\left({\rho_p\over\rho_n}\right)_p\delta\beta^*\delta\beta\nonumber\\
&&+\left({\partial\rho\over\partial\beta}\right)_p
{\rm Re}[(2\delta U +\delta\Phi)\delta\beta^*]\biggr\}d^{\,3}x.
\label{6.1}
\eeqa

\noindent The integrals of the first two terms on the right side of
Eq.~(\ref{6.1}) are to be performed throughout the star, while the
last three terms (those proportional to $\delta w^a$ or $\delta
\beta$) are to be integrated only within the superfluid core.  In the
absence of dissipation, the energy ${\cal E}$ is conserved.
When we evaluate the small angular-velocity expansion
of the energy in Eq.~(\ref{6.1}), we find that only the first term on
the right contributes at the lowest order.  Thus, the lowest-order
expression for the energy,

\beqa
{\cal E}=&&\alpha^2 {\pi(m+1)^3\over 2m}(2m+1)!\times\nonumber\\ 
&&\quad R_0^{\,4}\Omega^2
\int_0^{R_0}\rho \left({r\over R_0}\right)^{2m+2} dr
+{\cal O}(\Omega^4),\label{6.2}
\eeqa

\noindent is identical to its ordinary-fluid counterpart\cite{lmo}.

Mutual friction tends to damp out any relative motion between the
proton and neutron superfluids.  This dissipation is caused by the
scattering of electrons (whose motion tracks that of the proton
superfluid in our approximation) off the magnetic fields that are
entrapped within the cores of the neutron vortices.  The presence of
these magnetic fields is due to the entrainment effect, and hence the
magnitude of this dissipation mechanism depends strongly on the poorly
known mass-matrix element $\rho_{np}$.  The rate at which energy is
dissipated by mutual friction in the neutron-star superfluid
considered here is given by~\cite{mendell1991b}:

\beqa \left({d{\cal E}\over dt}\right)_{\scriptscriptstyle MF}
 = &&-2\Omega\int B_n \rho_n
\gamma^2(\delta^{ab} -z^az^b)\delta w_a\delta w^*_b d^{\,3}x, 
\nonumber \\&&\label{6.3}
\eeqa

\noindent where the dimensionless mutual friction scattering
coefficient, $B_n$, is given by,

\beqa
B_n\approx {\epsilon^2 \rho_p^{1/6}\over 1.96\times 10^{4}} 
&&\left({\rho \over \rho_p} - 1 \right)
\left(1-\epsilon+\epsilon{\rho\over\rho_p}\right)^{-3/2}.
\label{6.4}
\eeqa

\noindent We note that unlike regular viscosity, the mutual friction
scattering coefficient does not depend on temperature (at least at the
lowest order).  This independence is due to the fact that mutual
friction involves the scattering of two different fluids in relative
motion.  Thus, even at zero temperature some electrons acquire
sufficient energy from their collisions with the vortices to scatter
into available energy states located above the Fermi surface.

In order to evaluate the effects of mutual friction on the superfluid
$r$-modes, it is necessary to re-express Eq.~(\ref{6.3}) in terms of
the eigenfunction $\delta \beta_2$ that determines the relative
superfluid motion.  To lowest order in the angular velocity then, we
find

\beqa
&&\left({d{\cal E}\over dt}\right)_{\scriptscriptstyle MF} =
-2R_0^{\,4} (\pi G \bar{\rho}_0)^{3/2}
\left({\Omega\over\sqrt{ \pi G \bar{\rho}_0}}\right)^{7}\nonumber\\
&&\qquad\times\int
{B_n\rho_0\gamma_0^2(1-\mu^2)\over(\kappa_0^2-4\gamma_0^2)^2}
\left(1-{\rho_p\over\rho}\right)_0\nonumber\\
&&\qquad\quad
\times\Biggl[\biggl|\kappa_0\biggl({\partial\delta\beta_2\over\partial r}
-{\mu\over r}{\partial\delta\beta_2\over\partial\mu}\biggr)
+{2m\gamma_0\delta\beta_2\over r(1-\mu^2)}\biggr|^2\nonumber\\
&&\qquad\qquad+\biggl|2\gamma_0\biggl({\partial\delta\beta_2\over\partial r}
-{\mu\over r}{\partial\delta\beta_2\over\partial\mu}\biggr)
+{m\kappa_0\delta\beta_2\over r(1-\mu^2)}\biggr|^2\Biggr]d^{\,3}x.
\nonumber\\
\label{6.5}
\eeqa

\noindent It is straightforward to perform these integrals numerically
using the superfluid $r$-mode eigenfunctions described in
Sec.~\ref{sectionV}, and so determine the energy dissipation rate
caused by mutual friction.  It is convenient to express this rate in
terms of a mutual-friction damping time, $\tau_{\scriptscriptstyle
MF}$:

\beq {1\over\tau_{\scriptscriptstyle MF}} 
=-{1\over 2{\cal E}} \left({d {\cal E}\over dt}\right)_{\scriptscriptstyle
MF}
= {1\over\tilde{\tau}_{\scriptscriptstyle MF}}
\left({\Omega\over\sqrt{ \pi G \bar{\rho}_0}}\right)^{5}.
\label{6.6}
\eeq

\bfig \centerline{\psfig{file=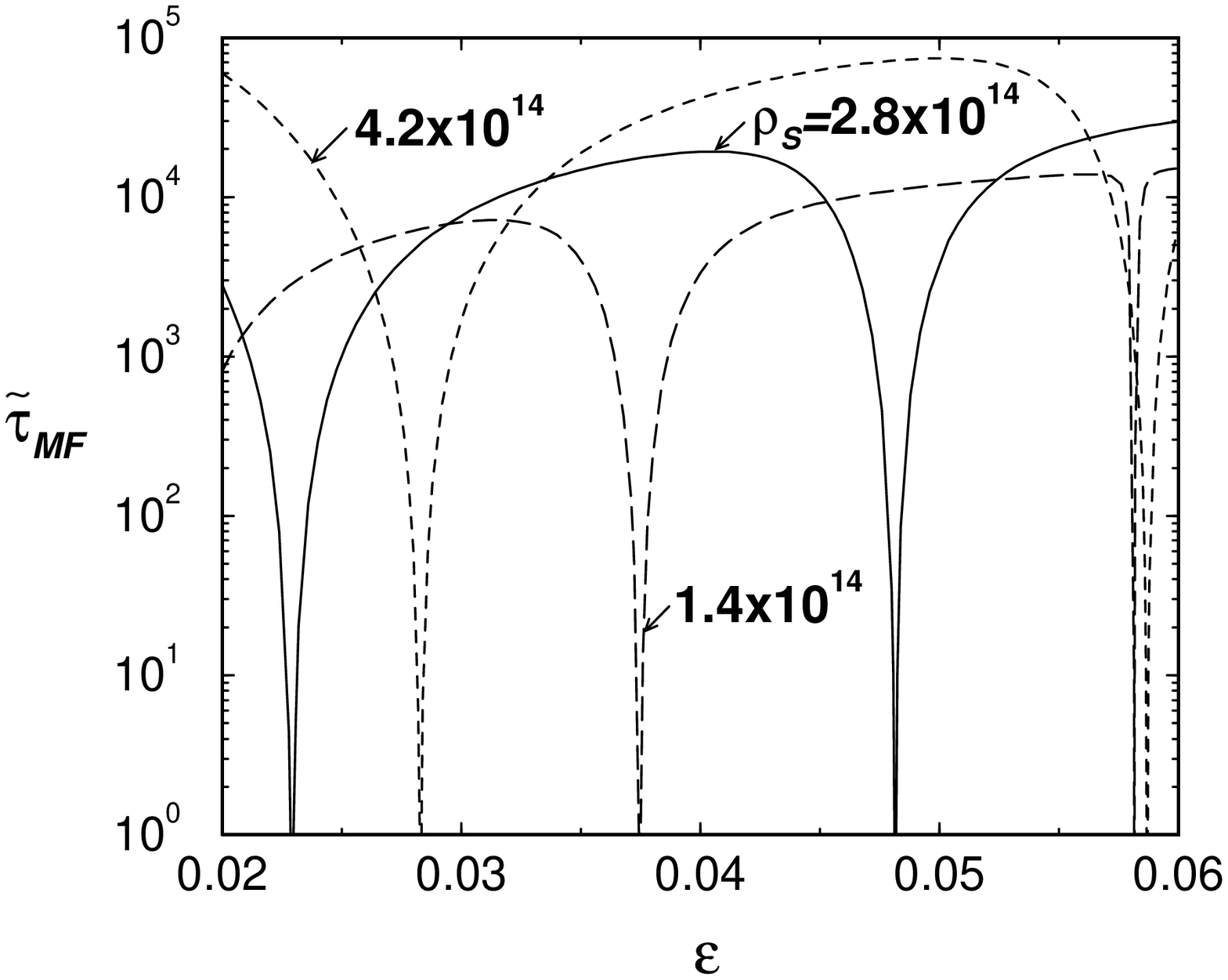,height=2.4in}} \vskip 0.3cm
\caption{Characteristic damping time $\tilde{\tau}_{\scriptscriptstyle MF}$
due to superfluid mutual friction.  These curves show the dependence
of $\tilde{\tau}_{\scriptscriptstyle MF}$ as a function of the superfluid
entrainment parameter $\epsilon$ for three values of the superfluid transition
density $\rho_s$.\label{fig6}} \efig

\bfig \centerline{\psfig{file=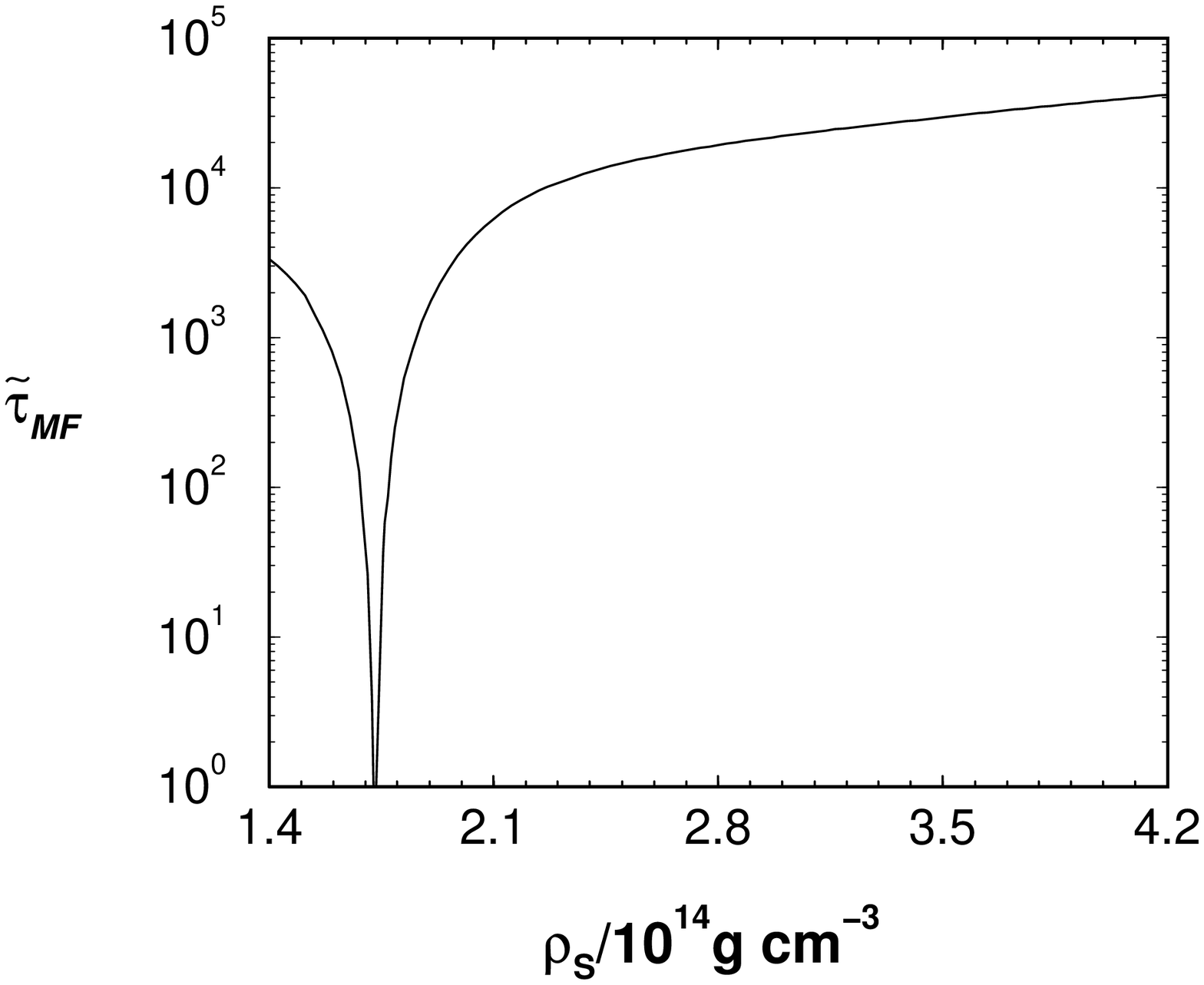,height=2.4in}} \vskip 0.3cm
\caption{Characteristic damping time $\tilde{\tau}_{\scriptscriptstyle
MF}$ due to superfluid mutual friction.  This curve shows the
dependence of $\tilde{\tau}_{\scriptscriptstyle MF}$ as a function of
the superfluid transition density $\rho_s$ for
$\epsilon=0.04$.\label{fig7}} \efig

\noindent The characteristic mutual-friction damping time
$\tilde{\tau}_{\scriptscriptstyle MF}$, also defined in
Eq.~(\ref{6.6}), is independent of angular velocity and temperature
(to lowest order).  The $\Omega^5$ scaling of
$1/\tau_{\scriptscriptstyle MF}$ follows directly from
Eqs.~(\ref{6.2}) and (\ref{6.5}).  The $\Omega^7$ scaling of $d{\cal
E}/dt$ in Eq.~(\ref{6.5}) follows in turn from Eq.~(\ref{6.3}) and the
fact that $\delta w^a$ scales as $\Omega^3$ to lowest order.  We
present in Figs.~\ref{fig6} and \ref{fig7} the numerically determined
characteristic damping times $\tilde{\tau}_{\scriptscriptstyle MF}$
for the $m=2$ superfluid $r$-modes.  Fig.~\ref{fig6} illustrates how
sensitively $\tilde{\tau}_{\scriptscriptstyle MF}$ depends on the
entrainment parameter $\epsilon$.  We see that
$\tilde{\tau}_{\scriptscriptstyle MF}$ has a typical value of about
$10^4$ s, but that it becomes {\it much} smaller for a few narrow
ranges of $\epsilon$.  These spikes in the
$\tilde{\tau}_{\scriptscriptstyle MF}(\epsilon)$ curves are caused by
the ``resonance'' phenomenon that we discuss in Sec.~\ref{sectionV}.
The mutual friction damping time $\tilde{\tau}_{\scriptscriptstyle
MF}$ becomes very small when the eigenfunction $\delta \beta_2$
becomes large.  This occurs at the points where the operator
$\tilde{D}$ has a vanishing eigenvalue.  We see from Fig.~\ref{fig5}
that the locations of these vanishing eigenvalues coincide exactly
with the location of the spikes in the $\rho_s=2.8\times
10^{14}$~g/cm${}^{3}$ curve in Fig.~\ref{fig6}.  We find that near
these spikes the curve $\tilde{\tau}_{\scriptscriptstyle
MF}(\epsilon)\propto (\epsilon-\epsilon_c)^2$.  This quadratic
dependence is exactly what is expected given that $\delta\beta_2
\propto \psi_{\min}/\lambda_{\min}$ near these spikes.
Fig.~\ref{fig6} also illustrates that the exact location of these
spikes depends on the value of the superfluid transition density
$\rho_s$.  Fig.~\ref{fig7} illustrates that
$\tilde{\tau}_{\scriptscriptstyle MF}$ changes smoothly as $\rho_s$ is
varied: the locations of these spikes move smoothly to larger values
of $\epsilon$ as $\rho_s$ increases.  Fig.~\ref{fig6} also illustrates
that the probability of having a small
$\tilde{\tau}_{\scriptscriptstyle MF}$ does not depend strongly on
$\rho_s$.  For example, we find that about 1\% of the values of
$\epsilon$ in the acceptable range, $0.02\leq\epsilon\leq0.06$, have
$\tilde{\tau}_{\scriptscriptstyle MF}(\epsilon)\leq 5$~s, and this
percentage is relatively insensitive to $\rho_s$.

The other important internal fluid dissipation mechanism in superfluid
neutron stars is expected to be regular shear viscosity.  In the
superfluid core this viscosity is due to electron-electron scattering,
while in the surrounding ordinary-fluid envelope the standard
neutron-neutron scattering dominates.  The rate at which energy is
dissipated by shear viscosity is given by:

\beqa \left({d{\cal E}\over dt}\right)_{\scriptscriptstyle V} =
&&-\int 2\eta \delta\sigma^*_{ab} \delta\sigma^{ab}d^{\,3}x
+ \int_s 2\eta n^a\delta v^b_e\delta\sigma^*_{ab}d^{\,2}x\nonumber
\\ &&\qquad
-\int_o 2\eta n^a\delta
v^b\delta\sigma^*_{ab}d^{\,2}x.  
\label{6.7}\eeqa

\noindent The volume integral in Eq.~(\ref{6.7}) is to be evaluated
within the superfluid core, and within the ordinary-fluid envelope,
but not over the boundary surface between the two.  The surface
integrals are to be evaluated over the interior (superfluid side) and
exterior (ordinary-fluid side) of the superfluid boundary
respectively.  The tensor $\delta\sigma_{ab}$ that appears in these
integrals is the shear of the electron velocity $\delta v^a_e$ of
Eq.~(\ref{2.4}) within the superfluid core, and the ordinary-fluid
velocity $\delta v^a$ in the envelope.  In the small angular velocity
expansion the electron velocity $\delta v_e^a$ is the same as $\delta
v^a$, to lowest order.  Thus to lowest order the energy dissipation
rate due to shear viscosity is given by

\beqa
\left({d{\cal E}\over dt}\right)_{\scriptscriptstyle V} &&=
-\alpha^2
{\pi(m+1)^3\over m}(m-1)(2m+1)!R_0^2\Omega^2\times\nonumber\\
&&\quad
\Biggl\{(2m+1)\int_0^{R_0}\eta\left({r\over R_0}\right)^{2m}dr\nonumber\\
&&\qquad\qquad\qquad
- (\eta_s-\eta_o)R_s
\left({R_s\over R_0}\right)^{2m}\Biggr\},
\label{6.8}
\eeqa

\noindent where $\eta_s$ and $\eta_o$ are the limits of the viscosity
taken from the superfluid and the ordinary-fluid side of the boundary
respectively.

In the superfluid core of the neutron star, $r\leq R_s$, the
appropriate viscosity to use in Eq.~(\ref{6.8}) is due to
electron-electron scattering.  This electron-electron scattering
viscosity is given approximately by the expression~\cite{cutler-lind}

\beq
\eta=6.0\times 10^6 \left({\rho\over T}\right)^2,\label{6.9}
\eeq

\noindent where all quantities are given in cgs units.  Similarly,
in the ordinary-fluid envelope, neutron-neutron scattering viscosity
dominates at the densities where most of the dissipation occurs.  This
neutron-neutron scattering viscosity is given approximately by the
expression~\cite{cutler-lind}

\beq
\eta=347 {\rho^{9/4}\over T^{2}}.\label{6.10}
\eeq

\noindent Using these expressions for the viscosity it is straightforward
to evaluate the energy dissipation rate for shear viscosity using
Eq.~(\ref{6.8}).  As in the case of mutual friction, it is convenient
to express the result as a viscous time-scale:

\beq
{1\over \tau_{\scriptscriptstyle V}} = 
-{1\over 2{\cal E}}\left({d{\cal E}\over dt}\right)_{\scriptscriptstyle V}
={1\over \tilde{\tau}_{\scriptscriptstyle V}} \left({10^9 {\rm K}\over T}
\right)^2.
\label{6.11}
\eeq

\noindent The characteristic viscous time-scale
$\tilde{\tau}_{\scriptscriptstyle V}$, also defined in
Eq.~(\ref{6.11}), is independent of the angular velocity and the
temperature of the neutron star.  We find that
$\tilde{\tau}_{\scriptscriptstyle V} = 1.01\times 10^8$~s for our
superfluid neutron-star model with $\rho_s=2.8\times
10^{14}$~g/cm${}^3$.  This value is somewhat shorter than that
obtained for hot neutron stars, $2.52\times 10^8$~s, by Lindblom,
Owen, and Morsink~\cite{lom}, and for cold neutron stars, $2.25\times
10^8$~s, by Andersson, Kokkotas, and Schutz~\cite{aks}.

Gravitational radiation is the final form of dissipation expected to
have a significant influence on the $r$-modes of superfluid neutron
stars.  Since the superfluid $r$-modes are identical to their
ordinary-fluid counterparts to lowest order in the angular velocity,
the gravitational radiation coupling is the same to this order.  The
characteristic gravitational radiation time-scale
$\tilde{\tau}_{\scriptscriptstyle GR}$ for the $m=2$ $r$-mode,

\beq
{1\over\tau_{\scriptscriptstyle GR}}= 
-{1\over 2{\cal E}}\left({d{\cal E}\over dt}\right)_{\scriptscriptstyle GR}
=-{1\over \tilde{\tau}_{\scriptscriptstyle GR}} \left({\Omega\over
\sqrt{\pi G \bar{\rho}_0}}\right)^6,
\label{6.12}
\eeq

\noindent therefore has the same value,
$\tilde{\tau}_{\scriptscriptstyle GR}=3.26$~s, as in the
ordinary-fluid case~\cite{lom}.

The effects of mutual friction, shear viscosity, and gravitational
radiation act together simultaneously to influence the evolution of
the superfluid $r$-modes.  Their combined effects on the evolution
of the energy of the mode are conveniently described by the
overall dissipative time-scale, $\tau$:

\beqa
{1\over\tau} =&& 
-{1\over 2{\cal E}}
\Biggl\{\left({d{\cal E}\over dt}\right)_{\scriptscriptstyle GR}
+\left({d{\cal E}\over dt}\right)_{\scriptscriptstyle MF}
+\left({d{\cal E}\over dt}\right)_{\scriptscriptstyle V}
\Biggr\}.\label{6.13}
\eeqa

\noindent Since the energy $\cal E$ is a real functional, the quantity
$1/\tau$ is in fact just the imaginary part of the frequency of the
mode.  The sign of $\tau$ therefore determines whether the mode is
stable. If $\tau>0$ then dissipation decreases the energy of the mode
and it is stable, while if $\tau<0$ then dissipation causes the energy
(and hence the mode itself) to increase exponentially. We can explore
the conditions under which mutual friction and viscosity are effective
in suppressing the gravitational radiation driven instability by
giving the explicit temperature and angular-velocity dependence of
$\tau$ using Eqs.~(\ref{6.6}), (\ref{6.11}), and (\ref{6.12}):

\beqa
{1\over\tau(\Omega,T)} = &&-{1\over\tilde{\tau}_{\scriptscriptstyle GR}}
\left({\Omega\over\sqrt{\pi G\bar{\rho}_0}}\right)^6
\nonumber\\
&&\quad\!+{1\over\tilde{\tau}_{\scriptscriptstyle MF}}
\left({\Omega\over\sqrt{\pi G\bar{\rho}_0}}\right)^5
+{1\over\tilde{\tau}_{\scriptscriptstyle V}}
\left({10^9{\rm K}\over T}\right)^2.
\label{6.14}
\eeqa

Gravitational radiation tends to drive the $r$-modes unstable while
mutual friction and shear viscosity tend to stabilize these modes.
From Eq.~(\ref{6.14}) we see that $\tau>0$, and mutual friction will
completely suppress the gravitational radiation driven instability
whenever

\beq
\tilde{\tau}_{\scriptscriptstyle MF}\left({\Omega\over\sqrt{\pi G
\bar{\rho}_0}}\right)\leq \tilde{\tau}_{\scriptscriptstyle GR}.
\label{6.15}
\eeq

\noindent Since neutron stars have angular velocities that are limited
by $\Omega\lesssim\case{2}{3}\sqrt{\pi G \bar{\rho}_0}$, we see that
mutual friction will suppress the gravitational radiation instability
for all angular velocities whenever $\tilde{\tau}_{\scriptscriptstyle
MF}\lesssim 1.5\tilde{\tau}_{\scriptscriptstyle GR}\approx 4.89$ s.
From Fig.~\ref{fig6} we see that this may occur, but only if the
entrainment parameter $\epsilon$ of neutron-star matter is limited to
a very narrow range.  Only about 1\% of the values of $\epsilon$ in
the expected range, $0.02\leq\epsilon\leq0.06$, have sufficiently
short $\tilde{\tau}_{\scriptscriptstyle MF}$ to suppress the
instability completely.

From Eq.~(\ref{6.15}) we see that mutual friction will always suppress
the gravitational radiation driven instability in the $r$-modes of
neutron stars with sufficiently small angular velocities.  However,
this suppression is not physically relevant if
$\tilde{\tau}_{\scriptscriptstyle MF}$ is too large.  If the angular
velocities needed in Eq.~(\ref{6.15}) are sufficiently small, then
shear viscosity will also play a significant role in suppressing the
$r$-mode instability in these stars.  In particular shear viscosity
alone will suppress the $r$-mode instability if

\beqa
{\Omega\over \sqrt{\pi G \bar{\rho}_0}} &&\leq 
\left({\tilde{\tau}_{\scriptscriptstyle GR}\over
\tilde{\tau}_{\scriptscriptstyle V}}\right)^{1/6}
\left({10^9{\rm K}\over T}\right)^{1/3}\approx 0.0564
\left({10^9{\rm K}\over T}\right)^{1/3}.\nonumber\\
\label{6.16}
\eeqa

\noindent Mutual friction then is capable of further suppressing the
$r$-mode instability in more rapidly rotating neutron stars only if

\beqa
\tilde{\tau}_{\scriptscriptstyle MF}&&\lesssim
\tilde{\tau}_{\scriptscriptstyle GR}
\left({\tilde{\tau}_{\scriptscriptstyle V}\over
\tilde{\tau}_{\scriptscriptstyle GR}}\right)^{1/6}
\left({T\over 10^9{\rm K}}\right)^{1/3}
\approx 57.7 {\rm s}\left({T\over 10^9{\rm K}}\right)^{1/3}.
\nonumber\\
\label{6.17}
\eeqa

\noindent We see that mutual friction will be primarily responsible
for the suppression of the $r$-mode instability in some (sufficiently
warm) superfluid neutron stars only if
$\tilde{\tau}_{\scriptscriptstyle MF}\lesssim57.7$~s.  Such small
values of $\tilde{\tau}_{\scriptscriptstyle MF}$ occur in only about
3\% of the values of $\epsilon$ in the expected range:
$0.02\leq\epsilon\leq0.06$.

It is also informative to examine the critical angular velocities,
$\Omega_c$, that mark the dividing line between stable and unstable
neutron stars: $\tau(\Omega_c,T)=0$.  Stars rotating more rapidly than
$\Omega_c$ are unstable, while those rotating more slowly are stable.
Fig.~\ref{fig8} illustrates the temperature dependence of $\Omega_c$
for a range of possible mutual friction time-scales.  From
Fig.~\ref{fig8} we see that shear viscosity completely suppresses the
gravitational radiation instability in all neutron stars cooler than
about $10^6$~K.  We also see that the mutual friction time-scale must
be shorter than 10 or 15 s in order for mutual friction to play a
significant role in suppressing the gravitational radiation
instability in neutron stars with temperatures that are typical of low
mass x-ray binaries (about $10^8$~K).  Since only about 2\% of the
expected range of $\epsilon$ have time-scales this short, it appears
unlikely that mutual friction is acting to suppress the gravitational
radiation instability in these stars.

\bfig \centerline{\psfig{file=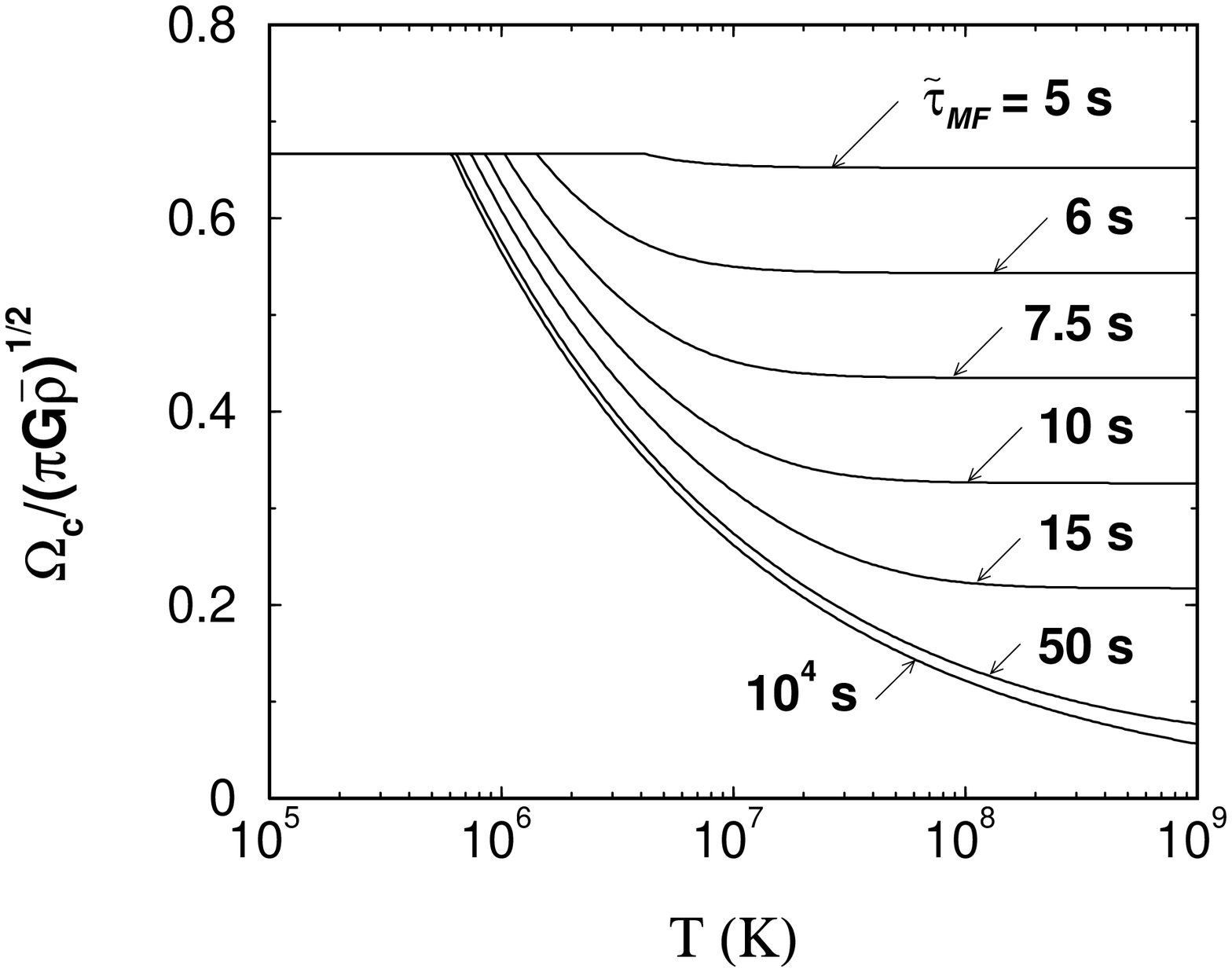,height=2.4in}} \vskip 0.3cm
\caption{Critical angular velocities for superfluid neutron stars
having a range of characteristic mutual friction time-scales
$\tilde{\tau}_{\scriptscriptstyle MF}$.\label{fig8}} \efig

\acknowledgments We thank L.~Bildsten, E.~Brown and Y.~Levin for
helpful discussions concerning this work.  This research was supported
by NSF grant PHY-9796079 and NASA grant NAG5-4093.

\appendix
\section*{Pulsation Equations in Spherical Coordinates}
\label{appendix}

In order to solve the superfluid pulsation equations numerically, it
is necessary to express them in some particular coordinate
representation.  We find it useful to work in spherical coordinate:
$r$, $\mu=\cos\theta$, and $\varphi$.  The operators $D$, $E$, $F$,
$\tilde{D}$, $\tilde{E}$ and $\tilde{F}$ that appear in
Eqs.~(\ref{4.18}) and (\ref{4.19}) have the following representations
in spherical coordinates,

\beqa
&&D(\delta U_2)
=\kappa_0^2\,\rho_0 \biggl[ {\partial^2\delta U_2\over\partial r^2}
+{1-\mu^2\over r^2}{\partial^2\delta U_2\over \partial\mu^2}
+{2\over r}{\partial\delta U_2\over \partial r}\nonumber \\
&&\qquad
-{2\mu\over r^2}{\partial\delta U_2\over\partial\mu}
-{m^2\delta U_2\over r^2(1-\mu^2)}\biggr]
-4\rho_0\biggl[
{2\mu(1-\mu^2)\over r}{\partial^2 \delta U_2\over \partial r\partial\mu}
\nonumber \\
&&\qquad
+\mu^2{\partial^2\delta U_2\over\partial r^2}
+{(1-\mu^2)^2\over r^2}{\partial^2\delta U_2\over \partial\mu^2}
+{1-\mu^2\over r}{\partial\delta U_2\over \partial r}
\nonumber \\
&&\qquad
-{3\mu(1-\mu^2)\over r^2}{\partial\delta U_2\over \partial\mu}\biggr]
+{d\rho_0\over dr}\biggl[
(\kappa_0^2-4\mu^2){\partial\delta U_2\over \partial r}
\nonumber \\
&&\qquad
+{2m\kappa_0\over r}\delta U_2
-{4\mu(1-\mu^2)\over r}{\partial\delta U_2\over \partial\mu}
\biggr],
\label{a1}
\eeqa

\beqa
&&E(\beta_2) =(\kappa_0^2-4\mu^2)r{d\over dr}
\biggl[{1\over r}{dh_0\over dr}
\left({\partial\rho\over\partial\beta}\right)_p
\biggr]\delta\beta_2\nonumber\\
&&\quad+ {dh_0\over dr}
\left({\partial\rho\over\partial\beta}\right)_p\biggl[(\kappa_0^2-4\mu^2)
{\partial\delta\beta_2\over\partial r}-{4\mu(1-\mu^2)\over r}
{\partial\delta\beta_2\over\partial\mu}\biggr]\nonumber\\
&&\quad+(3\kappa_0^2 - 2m\kappa_0 -4){1\over r}{dh_0\over dr}
\left({\partial \rho\over\partial\beta}\right)_p\delta\beta_2,
\label{a2}
\eeqa

\beqa
F =
&&{12m(m+2)\over (m+1)^2}{\rho_{22}\over r^2} \delta U_0
-2(m+2)\kappa_2{1\over r}{d\rho_0\over dr} \delta U_0\nonumber \\
&&+16\pi G\bar{\rho}_0 {m(m+2)\over (m+1)^4}
\left({d\rho\over dh}\right)_0(\delta U_0+\delta \Phi_0),
\label{a3}
\eeqa
\vfill\break
\beqa
&&\tilde{D}(\delta \beta_2)
={\kappa_0^2\,\tilde{\rho}_0\over\kappa_0^2-4\gamma_0^2}
\biggl[ {\partial^2\delta \beta_2\over\partial r^2}
+{1-\mu^2\over r^2}{\partial^2\delta \beta_2\over \partial\mu^2}
+{2\over r}{\partial\delta \beta_2\over \partial r}\nonumber \\
&&\quad
-{2\mu\over r^2}{\partial\delta \beta_2\over\partial\mu}
-{m^2\delta \beta_2\over r^2(1-\mu^2)}\biggr]
-{4\gamma_0^2\,\tilde{\rho}_0\over\kappa_0^2-4\gamma_0^2}
\biggl[\mu^2{\partial^2\delta \beta_2\over\partial r^2}
\nonumber \\
&&\quad
+{2\mu(1-\mu^2)\over r}{\partial^2 \delta \beta_2\over \partial
r\partial\mu}
+{(1-\mu^2)^2\over r^2}{\partial^2\delta \beta_2\over \partial\mu^2}
+{1-\mu^2\over r}{\partial\delta \beta_2\over \partial r}
\nonumber \\
&&\quad
-{3\mu(1-\mu^2)\over r^2}{\partial\delta \beta_2\over \partial\mu}\biggr]
+{d\over dr}\biggl({\kappa_0^2\,\tilde{\rho}_0\over \kappa_0^2-4\gamma_0^2}
\biggr){\partial\delta\beta_2\over\partial r}
\nonumber \\
&&\quad
-{d\over dr}\biggl({4\gamma_0^2\,\tilde{\rho}_0\over \kappa_0^2-4\gamma_0^2}
\biggr)\biggl[\mu^2{\partial\delta\beta_2\over\partial r}
+{\mu(1-\mu^2)\over r}{\partial\delta\beta_2\over\partial\mu}\biggr]
\nonumber\\
&&\quad
-{\kappa_0^2-4\mu^2\over\rho_0(\kappa_0^2-4)}
\left({\partial\rho\over\partial\beta}\right)^2_p
\left({dh_0\over dr}\right)^2\delta\beta_2
\nonumber\\
&&\quad+{2m\kappa_0\over r}{d\over dr}\biggl(
{\gamma_0\,\tilde{\rho}_0\over \kappa_0^2-4\gamma_0^2}\biggr)
\delta\beta_2,
\label{a4}
\eeqa

\beqa
&&\tilde{E}(\delta U_2) = - {1\over \kappa_0^2 -4}
{dh_0\over dr}\left({\partial\rho\over\partial\beta}\right)_p
\biggl[(\kappa_0^2-4\mu^2){\partial\delta U_2\over\partial r}
\nonumber \\&&\qquad\qquad\qquad
-{4\mu(1-\mu^2)\over r} {\partial \delta U_2\over \partial\mu}
+ {2m\kappa_0\over r}\delta U_2\biggr],
\label{a5}
\eeqa

\beqa
\tilde{F}=&&\left({\partial\rho\over\partial\beta}\right)_p
\left[{(m+1)^2\kappa_2\over 2mr}{dh_0\over dr} - {3h_{22}\over r^2}
\right]\delta U_0\nonumber\\
&&\qquad\qquad-{4\pi G \bar{\rho}_0\over (m+1)^2}
\left({\partial\rho\over\partial\beta}\right)_p
\Bigl(\delta U_0+\delta\Phi_0\Bigr).
\label{a6}
\eeqa

\noindent We note that the functions $\rho_{22}$ and $h_{22}$ that
appear on the right sides of Eqs.~(\ref{a3}) and (\ref{a6}) are parts
of the second-order expansions of these quantities.  In general the
second-order density function has the form:
$\rho_{2}(r,\mu)=\rho_{20}(r) + \rho_{22}(r)P_2(\mu)$, where
$\rho_{20}(r)$ and $\rho_{22}(r)$ are functions of $r$ and $P_2(\mu)$
is the $l=2$ Legendre polynomial.  A similar expression is satisfied
by $h_2$.  These functions may be determined using the techniques
described in Lindblom, Mendell, and Owen~\cite{lmo}.


\end{document}